%% file: main_acm.tex
\documentclass[sigconf,9pt]{acmart}

\usepackage{colortbl}
\usepackage{tikz}
\usepackage{enumitem}
\usepackage{amsmath,amssymb,amsfonts}
\usepackage{verbatim}
\usepackage{float}
\usepackage{caption}
\usepackage{amsmath,amsthm,amsfonts,amssymb,amscd,bm,amsbsy}
\usepackage{balance}
\usepackage{subfigure}
\usepackage{enumerate}
\usepackage{fancyhdr}
\usepackage{pifont}
\usepackage{mathrsfs}
\usepackage{xcolor}
\usepackage{graphicx}
\usepackage{listings}
\usepackage{xcolor}
\usepackage{soul}
\usepackage{capt-of}
\usepackage[normalem]{ulem}

\newcommand{\rev}[1]{#1}

\lstdefinestyle{sigcommcode}{
  basicstyle=\ttfamily\footnotesize,
  columns=fullflexible,
  breaklines=true,
  breakatwhitespace=true,
  frame=single,
  framerule=0.2pt,
  rulecolor=\color{black!30},
  backgroundcolor=\color{black!2},
  xleftmargin=1.2em,
  xrightmargin=0.6em,
  aboveskip=0.6em,
  belowskip=0.6em,
  showstringspaces=false,
  keepspaces=true,
  tabsize=2
}

\usepackage{float}
\usepackage{textcomp}
\usepackage{xcolor}
\usepackage{hyperref}
\usepackage{ragged2e}
\usepackage[utf8]{inputenc}
\usepackage{graphicx}
\usepackage{lipsum}
\usepackage{indentfirst}
\usepackage{hyperref}
\usepackage{cleveref}
\usepackage{tabularx}
\usepackage{balance}
\usepackage{wrapfig}
\usepackage{multirow}
\usepackage{capt-of}
\makeatletter 
\usepackage[ruled, vlined, linesnumbered]{algorithm2e}
\usepackage{algpseudocode}
\usepackage{tikz}
\setlist[enumerate]{label=(\arabic*)}
\usetikzlibrary{positioning, shapes, arrows.meta, fit, backgrounds}
\definecolor{gjr}{HTML}{FF0000}

\newcommand{\ourname}{Gryphon}

\acmYear{2026}\copyrightyear{2026}
\setcopyright{cc}
\setcctype[4.0]{by}
\acmConference[SIGCOMM '26]{ACM SIGCOMM 2026 Conference}{August 17--21, 2026}{Denver, CO, USA}
\acmBooktitle{ACM SIGCOMM 2026 Conference (SIGCOMM '26), August 17--21, 2026, Denver, CO, USA}
\acmDOI{10.1145/3789240.3829119}
\acmISBN{979-8-4007-2467-1/26/08}

\title[\ourname{}: Scaling Hyperscale Multi-Tenant Gateways]{\ourname{}: Scaling Hyperscale Multi-Tenant Gateways Beyond the Petabit-Era via DPU-Augmented Hierarchical Co-Offloading}

\newcommand{\pkuaff}{%
  \affiliation{%
    \institution{State Key Laboratory of Multimedia Information Processing,
    School of Computer Science, Peking University}
    \city{Beijing}
    \country{China}
  }%
}

\newcommand{\bdaff}{%
  \affiliation{%
    \institution{ByteDance}
    \city{Beijing}
    \country{China}
  }%
}

\author{Yuemeng Xu}
\pkuaff

\author{Haoran Chen}
\bdaff

\author{Jiarui Guo}
\pkuaff

\author{Mingwei Cui}
\bdaff

\author{Qiuheng Yin}
\pkuaff

\author{Cheng Dong}
\bdaff

\author{Peng He}
\bdaff

\author{Chenmin Sun}
\bdaff

\author{Yangyujia Wang}
\pkuaff

\author{Daxiang Kang}
\bdaff

\author{Xian Wu}
\bdaff

\author{Yang Gao}
\bdaff

\author{Lirong Lai}
\bdaff

\author{Kai Wang}
\bdaff

\author{Zhuochen Fan}
\pkuaff

\author{Tong Yang}
\pkuaff

\author{Hongyu Wu}
\bdaff

\makeatletter
\def\@mkauthors{%
  \begingroup
  \hsize=\textwidth
  \global\setbox\mktitle@bx=\vbox{%
    \noindent\unvbox\mktitle@bx\par\vskip 0.5em
    \centering
    {\large
    Yuemeng Xu$^{\dagger}$,
    Haoran Chen$^{*}$,
    Jiarui Guo$^{\dagger}$,
    Mingwei Cui$^{*}$,
    Qiuheng Yin$^{\dagger}$,
    Cheng Dong$^{*}$,\par
    Peng He$^{*}$,
    Chenmin Sun$^{*}$,
    Yangyujia Wang$^{\dagger}$,
    Daxiang Kang$^{*}$,
    Xian Wu$^{*}$,
    Yang Gao$^{*}$,\par
    Lirong Lai$^{*}$,
    Kai Wang$^{*}$,
    Zhuochen Fan$^{\dagger}$,
    Tong Yang$^{\dagger}$,
    Hongyu Wu$^{*}$\par
    }
    \vskip 0.35em
    {\normalsize
    $^{\dagger}$State Key Laboratory of Multimedia Information Processing,
    School of Computer Science, Peking University
    \quad
    $^{*}$ByteDance
    \par}
    \vskip 0.5em
  }%
  \endgroup
}
\makeatother

\renewcommand{\shortauthors}{Yuemeng Xu et al.}

\begin{document}

\input{content/new_abstract}



\maketitle
\sloppy

\input{content/intro_update914}
\input{content/background_motivation}
\input{content/Overview}

\input{content/design}

\input{content/Controlplane}
\input{content/deployment}
\input{content/implementation+}
\input{content/implementation_new}

\input{content/relatedwork}

\input{content/experience}

\input{content/Conclusion}

\clearpage

\begingroup
\enlargethispage{-\baselineskip}
\bibliographystyle{unsrt}
\bibliography{main}
\endgroup

\clearpage
\onecolumn
\section*{APPENDIX}
Appendices are supporting material that has not been peer-reviewed.
\par\medskip

\appendix
\input{content/appendix}

\end{document}

%% file: content/new_abstract.tex
\begin{abstract}

At ByteDance, cloud gateway clusters orchestrate petabit-scale aggregate traffic. 
Traditional ASIC-only gateways fail to meet these escalating demands due to severe on-chip resource constraints and limited programmable flexibility, while pure software solutions or alternatives like disaggregated SmartNICs struggle to match terabit-scale line-rate throughput. 
To bridge this gap, we present \ourname{}, a hyperscale cloud gateway built on a hybrid architecture that integrates DPUs directly into the switching ASIC’s forwarding path. 
This design resolves the fundamental tension between capacity and speed, expanding table scale by up to 1000$\times$ and augmenting programmability, while sustaining 1.6 Tbps line-rate throughput at a cost of only \rev{$\sim$}8\,$\mu\mathrm{s}$ in additional average latency.
To manage this hardware heterogeneity, we introduce Hierarchical Co-Offloading (HLCO) in the data plane, achieving >99.9\% fast path hit rate, while retaining software fallback for complex operations. 
In the control plane, we develop an abstraction layer (P4Bridge) that decouples hardware specifics from policy configuration. 
\ourname{} has been operating at production scale for over a year, deployed on hundreds of nodes across multiple Availability Zones. 
We also share production measurements and operational experiences that serve as the first hyperscale-proven guidelines for next-generation DPU-augmented cloud gateways.

\renewcommand{\thefootnote}{}
\footnotetext{The first three authors contribute equally. Tong Yang (yangtong@pku.edu.cn) and Hongyu Wu (hongyu.wu@bytedance.com) are corresponding authors.}
\renewcommand{\thefootnote}{\arabic{footnote}}

\end{abstract}

%% file: content/intro_update914.tex
\vspace{-0.05in}
\section{Introduction}

Managing petabit-scale aggregate traffic while maintaining strict requirements for tens of thousands of tenants presents unprecedented challenges in scale and complexity for centralized cloud gateways. 
Situated at the boundaries of datacenters, these gateways mediate a massive mix of intra-datacenter (east-west) and external (north-south) traffic~\cite{google:jupiter,google:maglev}, maintaining extensive dataplane entries for workloads ranging from throughput-intensive geo-distributed data synchronization~\cite{volcengine} to latency-sensitive interactive services~\cite{tiktok}. 
To sustain this, a single gateway instance must deliver microsecond-level latency, terabit-scale throughput, and accommodate hyperscale forwarding tables while remaining cost-efficient at scale~\cite{lu2025albatross,pan2024luoshen}. 

However, as cloud workloads evolve, the explosive growth of forwarding and policy rules, coupled with the requirement for strict isolation, has pushed commodity hardware architectures to their limits.
This strain manifests in three representative production workloads: 
\ding{172} Hyperscale tables (Doubao~\cite{doubao}): To strictly isolate thousands of tenants and services on our LLM platform, the control plane installs massive fine-grained forwarding rules and access-control lists (ACLs), quickly exhausting the limited on-chip memory (SRAM/TCAM) of commodity programmable switches~\cite{DBLP:journals/ccr/SivaramanMPNA20}. 
\ding{173} Low latency (TikTok~\cite{tiktok}): Interactive live streams require real-time server-side processing within cloud compute pools. 
Video flows are ingested at the edge and routed to the cloud core, demanding tight jitter and tail-latency bounds under contention~\cite{DBLP:journals/csur/DaoTTTBC22}. 
\ding{174} High throughput (Volcengine~\cite{volcengine}): Geo-redundant storage replication generates long-lived elephant flows that stress the gateway dataplane and can saturate its links, requiring high-throughput dataplane processing.

These diverse and demanding workloads expose the inherent limitations of existing gateway paradigms:
Software gateways (CPU-based) offer high flexibility but fail to scale with petabit-scale aggregate traffic due to packet processing bottlenecks~\cite{lei2024enhancing}. 
Conversely, while programmable switching ASICs (e.g., Tofino) deliver terabit-scale line-rate forwarding, they are constrained by severe on-chip memory limits, which remain insufficient to accommodate hyperscale forwarding tables even with algorithmic optimizations~\cite{pan2024luoshen}. 
Alternatives such as FPGA-based gateways, while providing greater flexibility than ASICs, incur substantial development complexity and significantly higher cost \cite{lu2025albatross}. 
This landscape reveals a critical gap: no existing architectures can simultaneously deliver the hyperscale capacity, deterministic performance, and rapid feature evolution required by modern cloud networks. 

A conventional approach to managing escalating workloads is horizontal scaling (scale-out) --- multiplying processing capability by clustering additional nodes. 
However, this strategy has become unsustainable for hyperscale gateways. 
For software-based clusters, a limited per-core packet-processing budget makes the system highly susceptible to hot-core (CPU hotspot) saturation under heavy-hitter–induced skew and microbursts~\cite{DBLP:conf/imc/ZhangLZK17,DBLP:conf/imc/FuscoD10,sigelman2010dapper}. 
Such single-core saturation inevitably compromises performance isolation among multiple tenants, leading to severe Service Level Agreement (SLA) violations~\cite{DBLP:conf/sc/ChenXZCWXSMYG20}. 
For ASIC-based solutions, scaling out to overcome memory limits necessitates fragmenting forwarding tables across multiple nodes, introducing immense complexity in inter-device traffic steering and state synchronization. 
Consequently, there is an urgent need for a scale-up architecture that enhances per-node capacity to sustain the demand of hyperscale workloads.

Integrating P4-programmable DPUs with established programmable switching ASICs offers a promising path to overcome these limitations. 
Modern DPUs feature massive DDR4/5 DRAM that breaks the on-chip SRAM capacity ceiling, and programmable MPUs capable of executing sophisticated logic beyond the fixed-stage constraints of ASIC pipelines.
This synergy creates a robust scale-up instance that simultaneously delivers hyperscale state capacity and terabit-scale line-rate throughput, all while incurring a latency overhead on the order of microseconds --- a trade-off fully acceptable for strict production SLAs~\cite{DBLP:conf/micro/GuoLBKSAL23}.

Realizing this heterogeneous architecture, however, is non-trivial and requires addressing three fundamental challenges:

\begin{itemize}[leftmargin=*, label=$\bullet$, itemsep=2pt, topsep=4pt]
    \item \textbf{Bridging the gap between host-centric DPUs and network-centric cloud gateways.}
    DPUs are primarily designed for per-host offloads, and typically provide 100–400 Gbps per device~\cite{kfoury2024smartnic}. 
    This is far below the Tbps-scale per device throughput demanded by hyperscale cloud gateways. Repurposing these endpoint devices for the network border introduces a fundamental performance mismatch~\cite{lu2025albatross}.
    \item \textbf{Orchestrating traffic across asymmetric hardware resources with extreme performance.}
    Switching ASICs deliver deterministic sub-microsecond match–action pipeline latency but are limited by scarce on-chip resources (SRAM/TCAM)~\cite{DBLP:conf/sigcomm/KimLZKLSS20,DBLP:journals/ccr/SivaramanMPNA20}. 
    In contrast, DPUs provide GB-scale forwarding state in off-chip DDR memory, but incur microsecond-scale latency~\cite{DBLP:conf/osdi/WeiCY0023} and low sustained throughput~\cite{DBLP:conf/micro/GuoLBKSAL23}. 
    Co-offloading traffic to harness these complementary strengths while mitigating their weaknesses is a significant challenge.
    \item \textbf{Shielding the control plane from hybrid datapath complexity.}
    Exposing raw hardware heterogeneity to upper layers impedes development velocity and hinders cross-team collaboration. 
    The challenge lies in designing a unified abstraction that enables operators to define logical policies agnostic to the underlying hardware, automatically mapping them to the appropriate target.
\end{itemize}

To address these challenges, we present \ourname{}, the first heterogeneous, hyperscale, multi-tenant cloud gateway that integrates programmable switching ASICs (Intel Tofino~\cite{DBLP:journals/ccr/BosshartDGIMRSTVVW14}) with P4-programmable DPUs (AMD Pensando~\cite{pensando2022}) into a single logical forwarding pipeline. 
We physically integrate four P4-programmable Pensando DPUs~\cite{pensando2022} with a folded Intel Tofino pipeline~\cite{pan2021sailfish}, aligning per-device bandwidth with gateway-scale throughput and achieving 1.6~Tbps line-rate forwarding. 
Centrally, we introduce Hierarchical Co-Offloading (HLCO), a tri-tier datapath that extends the classic fast/slow-path model~\cite{li2024triton}: 
the switching ASIC executes high-throughput, low-complexity processing, the DPU forms a flexible middle tier to host large tables and richer functions, and software provides a fully programmable fallback for exceptional or complex tasks. 
Finally, we design P4Bridge, a unified abstraction that exposes the ASIC-DPU datapath as a single logical pipeline, enabling transparent rule configuration across heterogeneous hardware. Our major contributions are summarized as follows: 

\begin{itemize}[leftmargin=1em]
  \item We present \ourname{}, the first cloud gateway that integrates DPUs and switching ASICs into a unified programmable P4 pipeline.
  It achieves 1.6 Tbps line-rate forwarding and leverages DPU memory to expand table capacity by $10\text{--}1000\times$, effectively eliminating resource hotspots on switching ASICs.
  \item We propose HLCO to efficiently partition tasks across the heterogeneous data plane. It offloads over 99.9\% of traffic to the hardware fast path, effectively eliminating software bottlenecks while maintaining software flexibility.
  \item We design P4Bridge, a hardware-agnostic framework that abstracts the hybrid ASIC-DPU datapath into a single logical pipeline. 
  We also introduce optimization techniques, including compact metadata encoding and table coalescing, to maximize interconnect efficiency and reduce memory access latency.
  \item We share experiences and lessons from a large-scale production deployment spanning over one year. 
  We analyze critical design trade-offs, providing the first hyperscale-proven guidelines for designing next-generation DPU-augmented cloud gateways.
\end{itemize}

\textbf{Ethics: This work does not raise any ethical issue.}

%% file: content/background_motivation.tex
\section{Background and Motivation}

\subsection{Anatomy of Cloud Gateways}

\begin{figure}[t] 
  \centering
  \includegraphics[width=.8\linewidth]{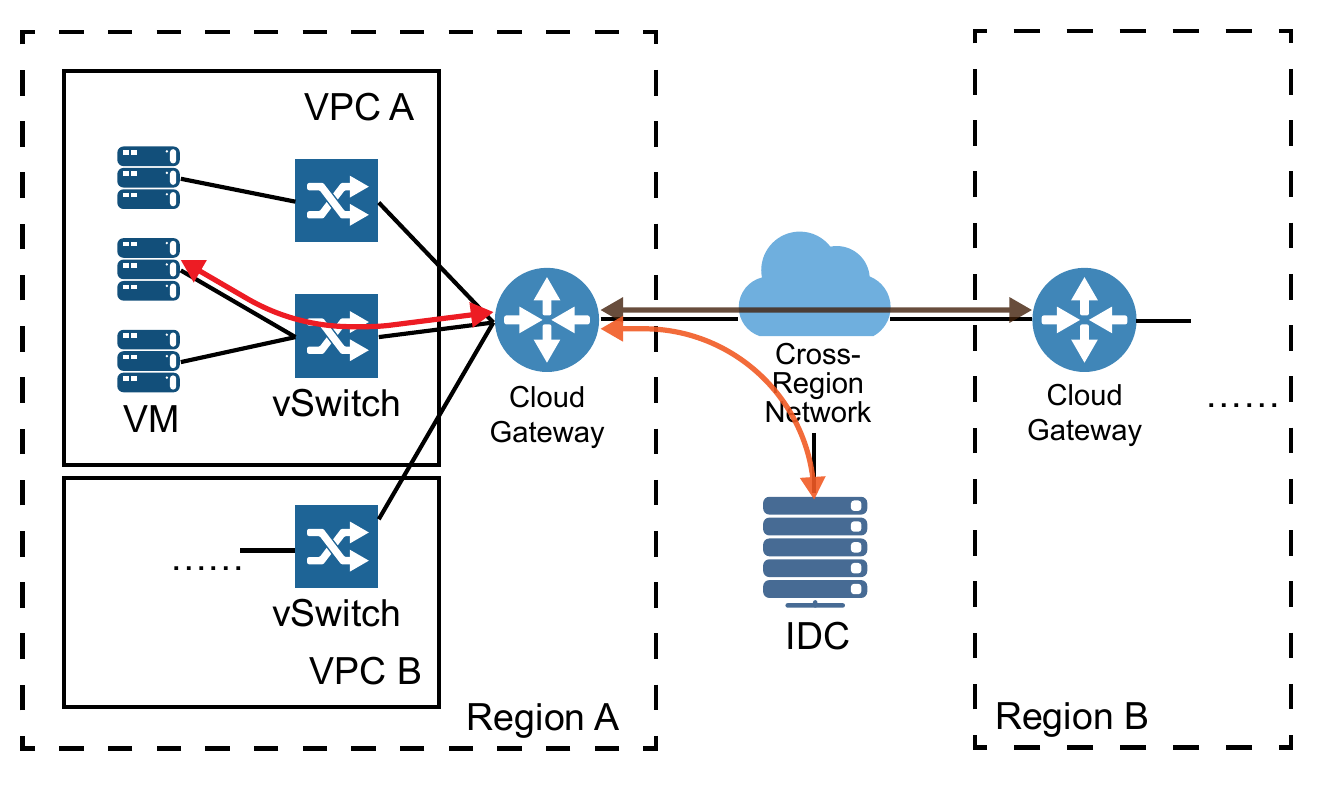}
  
  \vspace{-0.15in} 
  
  \caption{Cloud gateway as the central hub of traffic.}
  \vspace{-0.3in}
  \label{fig:topology}
\end{figure}

\textbf{Roles of cloud gateway.} 
In public clouds, providers multiplex a shared physical infrastructure into tenant-isolated virtual networks (e.g., VPCs). 
Each tenant operates within a private network abstraction with independent addressing, routing, and ACLs, requiring the underlying network to preserve strong isolation and fulfill SLA requirements \cite{DBLP:journals/ccr/SivaramanMPNA20,pan2024luoshen}. 
As illustrated in Figure~\ref{fig:topology} and Table~\ref{tab:connectivity}, the cloud gateway (CGW) serves as the central hub at critical traffic aggregation points: 
it orchestrates various connectivity patterns, connecting in-cloud VMs through the vSwitches, peering with remote CGWs and on-premises datacenters (IDC) over the cross-region network (CRN) \cite{pan2024luoshen}.
Operating at this junction, CGWs must sustain hyperscale throughput while enforcing fine-grained, multi-tenant forwarding policies. 



%

\textbf{Characteristics of hyperscale multi-tenant cloud gateway.} 
\ding{172} A significant portion of traffic is \emph{highly latency-sensitive} \cite{barker2010empirical, wu2023optimizing}. 
Critical applications, such as LLM inference, real-time collaboration, impose strict latency limits, where even minor delays can degrade service quality and compromise service level objectives (SLOs)~\cite{DBLP:journals/cacm/DeanB13}.
\ding{173} Traffic is characterized by its \emph{immense aggregate volume}, reaching the petabit-per-second (Pbps) level as a direct consequence of simultaneously serving numerous bandwidth-hungry workloads~\cite{google:jupiter,tiktok}.
\ding{174} The gateway must maintain \emph{massive per-tenant states}, including VPC routing, ACLs and VM-NC mappings, to enforce tenant isolation.
This results in an inevitable explosion of forwarding tables at the hyperscale traffic hub~\cite{pan2021sailfish,pan2024luoshen}.
\ding{175} Traffic exhibits a \emph{highly skewed flow size distribution} \cite{pasham2022graph}, characterized by long-lived, high-volume elephant flows alongside a vast number of short-lived mice flows.

\textbf{Design goals of hyperscale multi-tenant cloud gateway:} 
\ding{172} \textbf{Stringent performance.} 
The gateway must deliver microsecond-level latency and terabit-scale throughput to satisfy SLAs across diverse, bandwidth-intensive cloud workloads~\cite{google:jupiter,google:maglev,pan2021sailfish}.
\ding{173} \textbf{Hyperscale table capacity.} 
The gateway must maintain massive routing and policy tables to ensure tenant isolation~\cite{lu2025albatross}. 
\ding{174} \textbf{Cost efficiency.} 
At petabit scale, minimizing Total Cost of Ownership (TCO) is important, which requires co-optimizing hardware procurement (CapEx) and power consumption (OpEx) through judicious architectural choices.
\ding{175} \textbf{Flexibility.} 
The gateway must support rapid, safe updates to forwarding policies and packet-processing functions to enable fine-grained programmability and incremental feature rollouts~\cite{lu2025albatross}.

\subsection{Evolution towards \ourname{}}

\textbf{The first generation: Software-only cloud gateways.} 
Our initial cloud gateway adopted a fully software-based architecture~\cite{Kohler:Click:SOSP1999} and parallelized packet processing~\cite{Dobrescu:RouteBricks:SOSP2009,Han:PacketShader:SIGCOMM2010}. 
Driven by the need for rapid feature evolution, this design leveraged mature user-space packet I/O ecosystems (e.g., DPDK~\cite{dpdk_on_dpu}, Netmap~\cite{Rizzo:netmap:ATC2012}) to accelerate development cycles.

However, as traffic grew, tenant isolation became critical. 
While fixed per-attachment rate caps (e.g., a few~Gbps in our deployment) helped contain noisy neighbors, it became ill-suited for high-volume east-west traffic~\cite{pan2024luoshen}, leading to unnecessary queuing and packet loss under bursts. 
More fundamentally, scalability was often limited by traffic skew: hash-based mechanisms, such as Equal-Cost Multi-Path (ECMP) and Receive-Side Scaling (RSS), pinned long-lived elephant flows to specific paths or cores, and hash collisions among these flows created persistent hotspots~\cite{Hedera-NSDI10,DBLP:conf/sigcomm/AlizadehEDVCFLMPYV14}. 
As a result, aggregate throughput was frequently bottlenecked by the most congested core, reducing the effectiveness of horizontal scaling under skewed workloads~\cite{Hedera-NSDI10}.

\begin{table}[t]
  \caption{Connectivity provided by the cloud gateway.} 
  \vspace{-0.15in}
  \label{tab:connectivity}
  \small
  \begin{tabular}{l p{0.45\linewidth}}
    \toprule
    \textbf{Connectivity goal} & \textbf{Connectivity chain} \\
    \midrule
    internal VMs &
    CGW $\leftrightarrow$ vSwitch $\leftrightarrow$ VM \\
    \addlinespace[2pt]
    remote CGWs (inter-region) &
     CGW $\leftrightarrow$ CRN $\leftrightarrow$ CGW \\
    \addlinespace[2pt]
    IDC (on-premises) &
    CGW $\leftrightarrow$ CRN $\leftrightarrow$ IDC \\
    \bottomrule
  \end{tabular}
  \vspace{-0.2in}
\end{table}

\textbf{The second generation: Tofino-based gateways.} To overcome software bottlenecks, we transitioned to Tofino-based gateways, which deliver 1.6 Tbps line-rate throughput while 
achieving microsecond-level forwarding latency. 

\label{tab:sram-bottleneck}



However, the Tofino ASIC suffered from critically scarce on-chip memory~\cite{lu2025albatross}, making it challenging to support large-scale table entries, particularly the VM-NC tables mapping virtual machines (VMs) to their corresponding physical server addresses. 
Even with pipeline-aware resource packing and IPv4/IPv6 key compression, SRAM utilization frequently approached saturation. 
Table~\ref{tab:sram-bottleneck} illustrates this bottleneck, showing SRAM depletion across twelve consecutive stages within a critical pipeline where dependencies forced the placement of massive VM-NC tables.
Furthermore, the rigid match-action pipeline lacked specialized engines for computationally intensive tasks --- such as advanced load balancing, flow logging, and encryption --- and could not support the elastic buffering required for stateful operations \cite{DBLP:conf/sigcomm/SivaramanCBKABV16,DBLP:conf/sigcomm/BosshartGKVMIMH13,DBLP:conf/nsdi/FirestonePMCDAA18}. 
Consequently, these tasks were frequently punted back to software, resulting in unpredictable performance and increased operational complexity.

\begin{table*}[t]
    \centering
    \footnotesize 
    \caption{SRAM Utilization Hotspot in the Most Critical Pipeline Stages. }
    \vspace{-0.15in}
    \label{tab:sram-bottleneck}
    \begin{tabular*}{\textwidth}{@{\extracolsep{\fill}} lcccccccccccc r}
        \toprule
        \textbf{Metric} & \textbf{S1} & \textbf{S2} & \textbf{S3} & \textbf{S4} & \textbf{S5} & \textbf{S6} & \textbf{S7} & \textbf{S8} & \textbf{S9} & \textbf{S10} & \textbf{S11} & \textbf{S12} & \textbf{Average} \\
        \midrule
        SRAM Util. & \textbf{100.0\%} & \textbf{100.0\%} & \textbf{100.0\%} & 97.5\% & 98.8\% & 98.8\% & 98.8\% & \textbf{100.0\%} & 98.8\% & \textbf{100.0\%} & 98.8\% & 77.5\% & \textbf{97.4\%} \\
        \bottomrule
    \end{tabular*}
    \vspace{-0.15in}
\end{table*}

\textbf{Towards the third generation: Hybrid programmable gateway based on heterogeneous hardware.} 
Prior generations reveal a fundamental gap between software flexibility and ASIC performance, necessitating a new hardware tier to bridge them.
This tier must operate in tandem with Tofino, and deliver high programmability, competitive performance, and sufficient memory for hyperscale tables. 
We evaluated FPGAs and DPUs as candidates, and selected DPUs for their practical advantages in large-scale cloud deployments:

\begin{itemize}[leftmargin=1em]
    \item 
    \textbf{Higher efficiency and lower TCO.}
    Our benchmarking conducted under identical 2U chassis (same form factor and 1.6 Tbps line-rate forwarding) shows that DPU-based gateways reduce power consumption by 21\% and hardware cost by 14\% compared to commercial FPGA-based designs~\cite{DBLP:conf/usenix/WangH00A22, DBLP:journals/corr/abs-2504-03653}.
    This significantly lowers Total Cost of Ownership (TCO) at scale.
    \item \textbf{Reduced development complexity.}
    DPUs leverage a standard Linux environment and mature networking frameworks (e.g., C/C++, P4, and DPDK), enabling rapid iteration~\cite{dpdk_on_dpu,nvidia_dpu_whitepaper}. 
    In contrast, FPGA-based designs typically require development at the register-transfer level (RTL) or through high-level synthesis (HLS).
    They also involve longer synthesis and verification cycles, which significantly increases engineering overhead ~\cite{xilinx_hdl,xilinx_smartnic,DBLP:conf/iccad/LiaoWLLLY25,DBLP:conf/date/ZylaLWH25}.
    \item \textbf{Architectural compatibility.}
    DPUs and P4 switches share a similar match-action abstraction. 
    While the ASIC provides deterministic, TCAM-assisted prefix forwarding, the DPU's large DRAM and general-purpose compute support massive exact-match states and complex logic. 
    This alignment facilitates a natural hierarchical design and simplifies functional migration between components ~\cite{kim2023exoplane}.
\end{itemize}

\begin{figure}[t]
  \centering
  \includegraphics[width=0.5\textwidth]{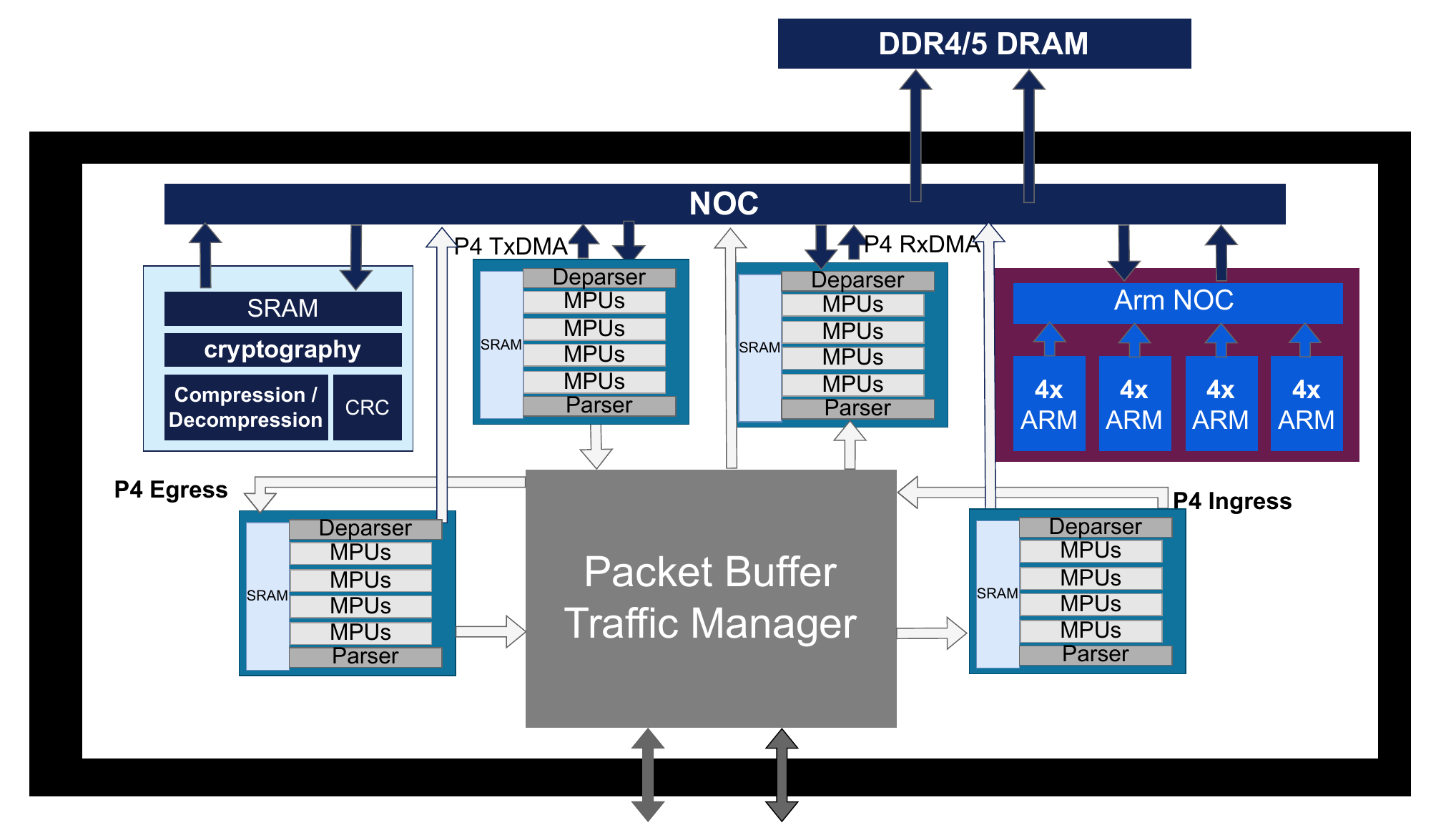} 
  \vspace{-0.35in}
  
 \caption{Simplified architecture of the DPU design.}

  \vspace{-0.35in}
  \label{fig:dpu_arch}
\end{figure}

\vspace{-0.05in}
\subsection{P4-programmable DPU}

To address the limitations of switching ASICs, we integrate P4-programmable DPUs into our next-generation gateway design. As shown in Figure~\ref{fig:dpu_arch}, the DPU provides a tightly coupled system-on-chip design, combining a programmable P4 dataplane with a general-purpose processing complex (ARM cores) that runs the DPDK-based datapath runtime.

The P4-programmed ingress and egress pipelines implement performance-critical dataplane functions such as access control (ACL), metering, routing, encapsulation/decapsulation, and lightweight NAT. To sustain large tables, the DPU employs on-chip SRAM as low-latency cache and off-chip DDR4/5 DRAM for high-capacity state storage.

Paired with dedicated Tx/Rx DMA engines, the DPU can achieve zero-copy data movement between P4 pipelines and system memory. This hybrid design bridges ASIC-class performance with software-defined flexibility, enabling scalable, low-latency packet processing and eliminating the need for host-side CPUs by running DPDK \cite{dpdk_on_dpu} runtime directly on embedded ARM cores.

%

%% file: content/Overview.tex
\section{\ourname{}'s Architecture}

\subsection{High-Level Architecture}
\label{subsec:high_level_architecture}
\begin{figure*}[t]
  \centering
  \begin{minipage}[t]{0.49\textwidth}
    \centering
    \vspace{1 cm}
    \subfigure[Logic overview of three typical datapaths for different types of traffic.]{%
      \includegraphics[width=\linewidth]{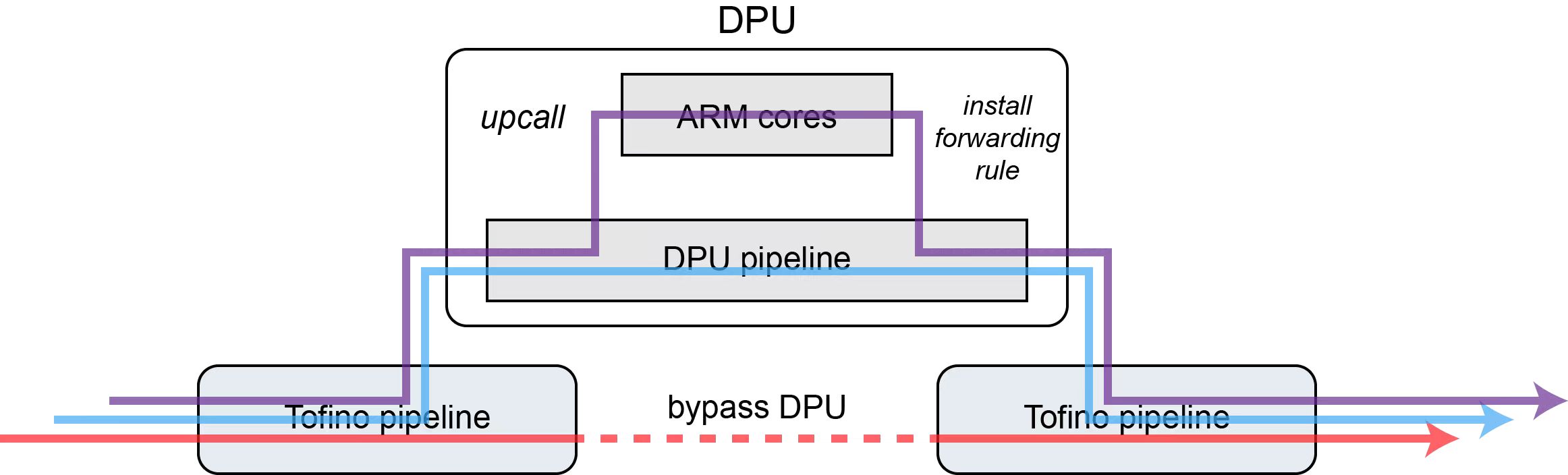}%
      \label{fig:pipeline}%
    }
  \end{minipage}\hfill
  \begin{minipage}[t]{0.49\textwidth}
    \centering
    \subfigure[Component view of the Tofino–DPU dataplane.]{%
      \includegraphics[width=\linewidth]{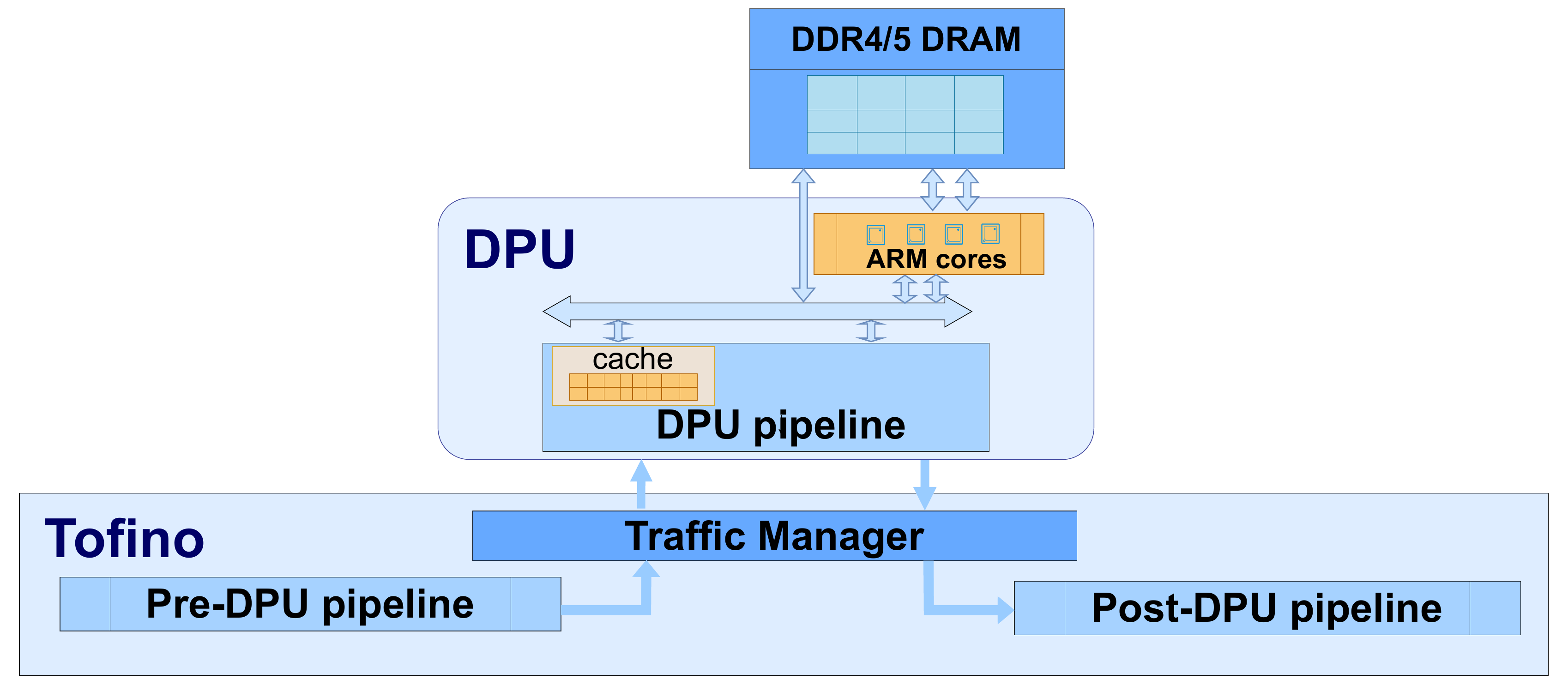}%
      \label{fig:macro_2}%
    }
  \end{minipage}
  \vspace{-0.15in}
  \caption{Architecture overview of hierarchically co-offloading design.}
  \vspace{-0.15in}
  \label{fig:arch_overview}
\end{figure*}

We extend the traditional fast/slow path model to a three-tier hierarchy: (1) switching ASICs as the ultra-fast path, (2) inline P4 pipelines on DPUs for mid-tier offload, and (3) ARM cores on DPUs \rev{handling} the ultimate fallback. This hierarchical offloading model preserves performance for the common case while retaining software flexibility.

Figures~\ref{fig:pipeline} and~\ref{fig:macro_2} summarize our hierarchically co-offloading design.
At a high level, the end-to-end packet-processing pipeline follows a three-stage datapath:
a pre-DPU Tofino pipeline, a DPU pipeline, and a post-DPU Tofino pipeline (Fig.~\ref{fig:pipeline}). Upon pipeline entry, each packet is tagged with a Service Identifier (SvcID) for classification and subsequent processing. All packets traverse the two Tofino stages, while the traffic manager steers only the packets that need additional state or computation to the DPU; the rest bypass the DPU stage and directly resume in the post-DPU pipeline. 

The DPU stage (Fig.~\ref{fig:macro_2}) integrates a programmable pipeline, ARM cores, and hierarchical memory (SRAM/DRAM). The pipeline performs exact-match lookups on DRAM-backed tables, utilizing on-chip SRAM as a low-latency cache. The pipeline accesses on-chip SRAM with deterministic latency, whereas external DRAM is managed via asynchronous lookup/DMA engines.

Upon a lookup hit (whether in SRAM or DRAM), the DPU returns the result to the post-DPU Tofino pipeline to complete the forwarding actions. Conversely, a lookup miss triggers an exception to the ARM cores for slow-path processing. The ARM cores resolve the forwarding decision and install the corresponding flow entries, ensuring subsequent packets are handled entirely in the hardware fast path.

\textbf{Path 1: Switch-only path.} This path handles traffic not destined for local VMs (e.g., cross-region, Internet, or IDC-bound). After traversing the pre-DPU Tofino pipeline, the packet is steered to bypass the DPU stage. It then proceeds directly to the post-DPU pipeline for forwarding, preserving the low-latency line-rate performance of the switching ASIC (Fig.~\ref{fig:pipeline}).

\textbf{Path 2: Hybrid fast path.} For traffic destined to local VMs, processing requires resolving the VM-NC table to identify the target physical host and determining the intra-datacenter next hop. However, the limited on-chip memory and processing stages of the Tofino ASIC are insufficient to accommodate these massive table entries. Consequently, such packets are forwarded to the DPU for exact-match lookups in SRAM/DRAM. Lookup hits return to the post-DPU pipeline where subsequent forwarding actions are executed (Fig.~\ref{fig:pipeline}); conversely, lookup misses are punted to the ARM cores.

\textbf{Path 3: Software slow path.} When the DPU pipeline fails to resolve a flow (i.e., a lookup miss)—typically triggered by the arrival of the first packet of a new flow—the packet is redirected to the ARM cores for software processing. The software agent executes the  control logic and installs corresponding forwarding entries, effectively offloading future traffic of the same flow to the Hybrid Fast Path (Path 2), bypassing the software stack.

\subsection{Hardware Organization}

Standard Tofino architectures operate four pipelines in parallel to maximize throughput, where each packet is processed by the ingress and egress logic within the same pipeline~\cite{DBLP:conf/sigcomm/BosshartGKVMIMH13}. However, for our logic-heavy cloud gateway, the primary bottleneck is not throughput, but the scarcity of processing stages and memory resources required to execute complex forwarding chains~\cite{pan2021sailfish}. Deployment data indicates that the stage and memory requirements of the most resource-intensive tables alone are sufficient to exhaust the resources of a standard pipeline. Furthermore, the integration of other functions (e.g., metering, ACL) introduces significant additional logic and memory footprints. 

Figure~\ref{fig:DPUAS} shows the detailed pipeline organization. We configure the internal ports between pipelines (pipes in Fig.~\ref{fig:DPUAS}) in a mode that chains the four pipelines sequentially. As illustrated, packets enter the first pipeline (pipe 3), traverse the ingress logic, and are redirected via the traffic manager to the egress logic of the subsequent pipeline (pipe 1), continuing this process until they reach the egress logic of the final pipeline (pipe 3) and exit the system. This "pipeline folding" to chain four physical pipelines into a single processing pipeline effectively multiplies the available Tofino resources by 4$\times$ to support rich forwarding logic, but at the cost of reducing aggregate throughput from 6.4 Tbps to 1.6 Tbps.

To match this line rate, we provision four DPUs in parallel and use ECMP to distribute packets to the DPUs, with packets entering and exiting the DPU via the same Tofino pipeline (pipe 2). This ``sandwich'' design co-locates the DPUs between the ingress and egress stages of the same Tofino pipeline (pipe 2), ensuring full-duplex bandwidth without sacrificing external port capacity.


%% file: content/design.tex
\begin{figure}[t]   
  \centering
  \includegraphics[width=0.8\linewidth]{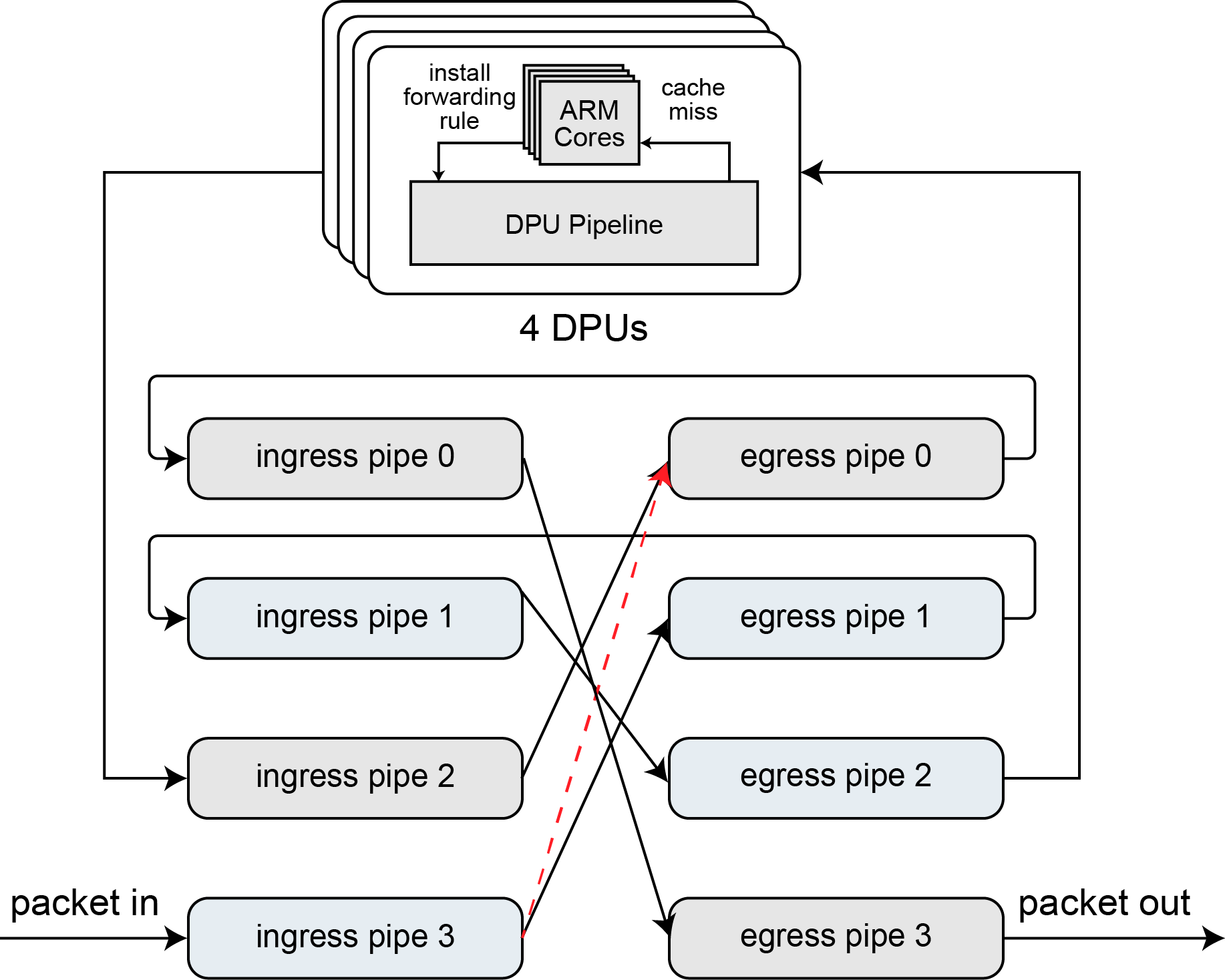}
  \caption{Datapath organization with folded pipeline and sandwiched DPUs.}
  \label{fig:DPUAS}
  \vspace{-0.3in}
\end{figure}

\begin{figure*}[t]   
  \centering
  \includegraphics[width=1.0\linewidth]{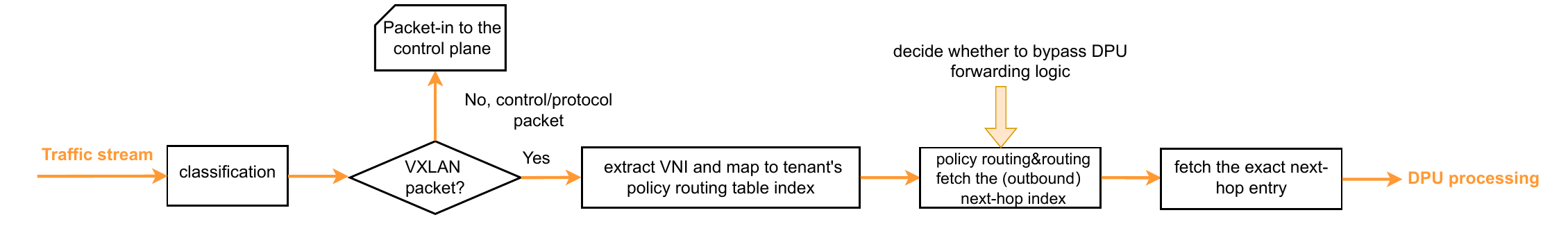}
  \vspace{-0.3in}
  \caption{Table lookup sequence in the pre-DPU Tofino pipeline.}
  \vspace{-0.15in}
  \label{fig:Tofin}
\end{figure*}

\section{Heterogeneous Pipeline}
\begin{figure*}[t]
	\centering
    \begin{minipage}[t]{0.5\textwidth}
        \includegraphics[width=1\textwidth]{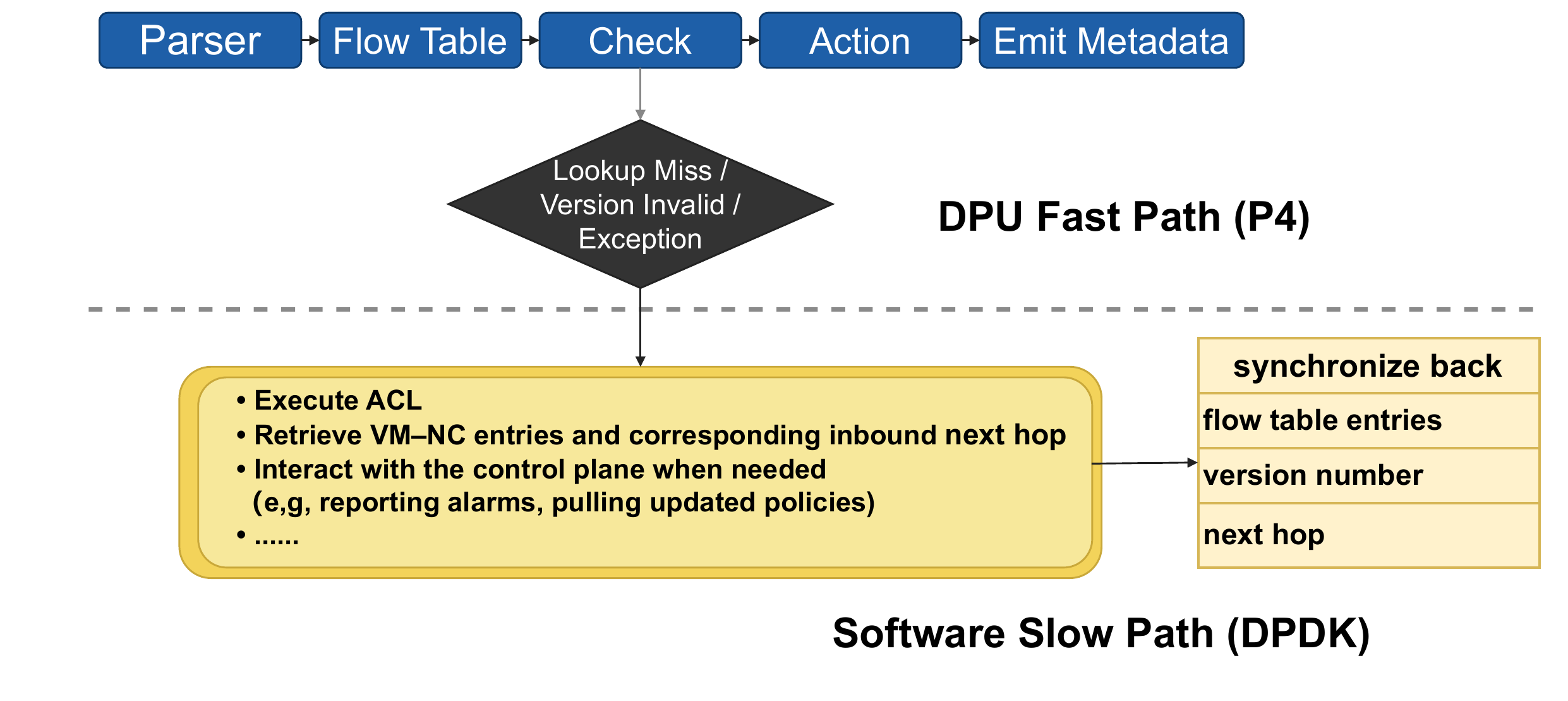}
        \vspace{-0.3in}
        \caption{DPU packet processing pipeline with fast/slow path interaction.}
        \vspace{-0.15in}
        \label{fig:DPUPP}
    \end{minipage}%
    \begin{minipage}[t]{0.5\textwidth}{
		\includegraphics[width=1\textwidth]{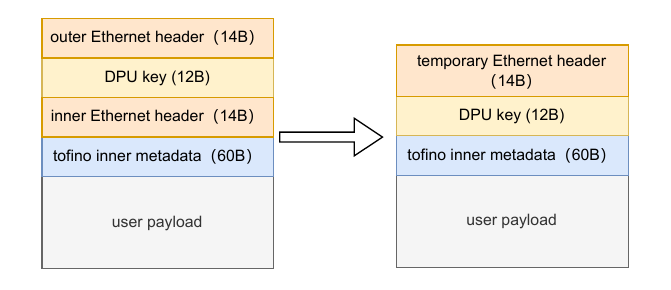}
        \vspace{-0.3in}
        \caption{Reducing extra overhead on the Tofino--DPU interconnect.}
        \vspace{-0.15in}
        \label{fig:meta_opt}
	}\end{minipage}
\end{figure*}

For a hyperscale cloud vendor, the datapath must support complex forwarding logic while adhering to strict hardware constraints. Instead of enumerating every corner case, we present a representative packet processing pipeline that illustrates the lifecycle of a typical VXLAN-encapsulated tenant packet across the heterogeneous \ourname{}'s architecture.

\textbf{Hardware-guided task partition.} Our design strategically maps different table types and processing logic across heterogeneous hardware based on their respective strengths. Leveraging the DPU's abundant SRAM and DRAM to overcome Tofino's on-chip memory scarcity, we offload hyperscale exact-match tables --- most notably the VM-NC table --- to the DPU. Conversely, the Tofino ASIC retains LPM lookups, utilizing its on-chip TCAM for deterministic prefix-based forwarding. Moreover, complex functions such as encryption, fine-grained flow logging, and ZooRoute~\cite{DBLP:conf/sigcomm/SunWLZY0YQYJZCL25} are executed on the DPU to leverage its specialized hardware accelerators and more flexible programmable pipelines. Finally, for lookup misses or policies requiring complex multi-step decisions, the DPU falls back to the software slow path, where the DPDK runtime (running on ARM cores) executes complex processing.

For protocol and control packets, DPU involvement is unnecessary. Allowing them to traverse the whole pipeline not only lengthens the processing path but also introduces potential risks to stability. To this end, we divert such traffic at ingress pipe~3 in Figure ~\ref{fig:DPUAS}, bypassing the DPU and delivering them directly to the CPU for control-plane processing as shown by the red path in Figure~\ref{fig:DPUAS}, ensuring that control-plane traffic follows the shortest and most stable path.

\textbf{Pre-DPU Tofino processing (ingress3 $\rightarrow$ egress1  $\rightarrow$ ingress1  $\rightarrow$ egress2 $\rightarrow$ DPU, in Figure ~\ref{fig:DPUAS}).} Pre-DPU processing follows three stages (detailed in Figure~\ref{fig:Tofin}):
(1) Stage 1: Traffic classification. Packets first hit a classification table that separates (a) VXLAN-encapsulated tenant traffic, identified by outer headers (e.g., UDP/VXLAN and remote VTEP IP), from (b) protocol/control traffic. Only VXLAN traffic proceeds through the subsequent stages.
(2) Stage 2: Tenant/context resolution. Next, a tenant identification table maps the VNI to a tenant context ID, which selects the corresponding (logically isolated) policy-routing namespace. This indirection is crucial for isolation: each tenant is bound to its own policy-routing instance/partition.
(3) Stage 3: Routing and forwarding. Within a tenant, multiple routing instances (e.g., subnets or security domains) may coexist; we select the routing instance using tenant context and a domain tag (e.g., derived from the inner source IP/subnet that associates source prefixes with routing domains or security contexts). Finally, a nexthop table materializes the forwarding action, including the output port and encapsulation parameters. Notably, the next hop applies only to outbound traffic not destined for local VMs, as the inbound next hop is resolved by the DPU.

\textbf{DPU Processing.} The DPU processes traffic destined for local VMs by leveraging its large table/memory capacity and flexible programmable logic. As illustrated in Figure~\ref{fig:DPUPP}, the DPU employs a hybrid architecture that integrates a P4-programmable hardware fast path with a DPDK-based software slow path. Upon entering the DPU's P4 ingress pipeline, packets undergo parsing and a flow table lookup to retrieve the nexthop index, the version number, and the destination physical host (NC) address resolved via the VM-NC mapping. In the event of a lookup miss or version validation failure, the packet is redirected to the software slow path. Within this path, the DPDK program handles lookup misses, expired rules, and complex corner cases. The DPDK runtime performs tasks including enforcing software-defined ACLs and resolving VM--NC mappings to determine the associated inbound nexthops. Additionally, it interacts with the control plane when needed to perform operations such as reporting runtime alarms. Once resolved, critical forwarding entries are synchronized back to the fast path, ensuring that subsequent packets from the same flow are handled entirely in hardware.

\textbf{Post-DPU Tofino processing (DPU $\rightarrow$ ingress2 $\rightarrow$ egress0  $\rightarrow$ ingress0  $\rightarrow$ egress3 in Figure ~\ref{fig:DPUAS}).}
After DPU processing, packets re-enter Tofino for a lightweight post-processing pass. Post-DPU Tofino pipeline typically (1) meters and enforces traffic budgets (e.g., per tenant), marking or dropping packets that exceed the configured rate; (2) handles packets that do not use VXLAN encapsulation by performing underlay Internet Protocol forwarding; (3) encapsulates overlay packets and attaches the selected underlay next-hop parameters; (4) recomputes and updates checksums after header rewrite or encapsulation to preserve packet correctness.

\textbf{Metadata design optimization.} Inter-hardware communication inevitably introduces header overhead, which reduces effective throughput because port bandwidth is bounded by the actual bytes transmitted on the link (especially under small packets)~\cite{pan2021sailfish}. To minimize this penalty, we treat the Tofino--DPU interconnect as a transient internal link and apply two optimization techniques.

First, we employ compact handle encoding: Tofino pre-resolves large-table references and exports only a compressed 1--3\,B index to the DPU. This yields a minimal 12\,B \textit{DPU key} (Figure~\ref{fig:meta_opt}, left).

Second, we implement header compression. \rev{In the uncompressed format, the
outer Ethernet header carries Tofino--DPU interconnect metadata, while the
inner one preserves the tenant header.} Rather than stacking both internal
and external Ethernet headers, we temporarily rewrite packets into a lean
transit format for the interconnect and restore the original headers upon
return (Figure~\ref{fig:meta_opt}, right). This saves 14\,B per packet.

Combined, these techniques limit the heterogeneous traversal overhead to just 12\,B, which translates into an effective-throughput gain on the order of 10\,Gb/s in our deployment.

%% file: content/Controlplane.tex
\section{Control Plane Synergy}
\subsection{Hardware Abstraction}

The vast architectural diversity among DPU vendors leads to distinct hardware architectures, programming abstractions, and resource constraints (e.g, varying capacities of TCAM, SRAM, or DRAM)~\cite{kfoury2024smartnic}. Moreover, although both Tofino ASIC and Pensando DPU adopt P4 as a common programming language, their underlying microarchitectural differences limit the portability of implementation expertise across platforms~\cite{DBLP:journals/ccr/BosshartDGIMRSTVVW14,kfoury2024smartnic}. Control-plane developers, tasked with orchestrating control-plane policies (e.g., ACLs, forwarding rules), often operate distinctly from the hardware engineering developers, necessitating a clear abstraction interface, as requiring control-plane operators to manage low-level data plane nuances (across diverse targets like ASICs and DPUs) would significantly increase complexity and impede development velocity. To address this fragmentation, we design an abstraction interface (P4Bridge) to decouple control plane logic from hardware-specific constraints and variations in data plane implementations.

\begin{figure}[t]
    \centering
    \includegraphics[width=.9\linewidth]{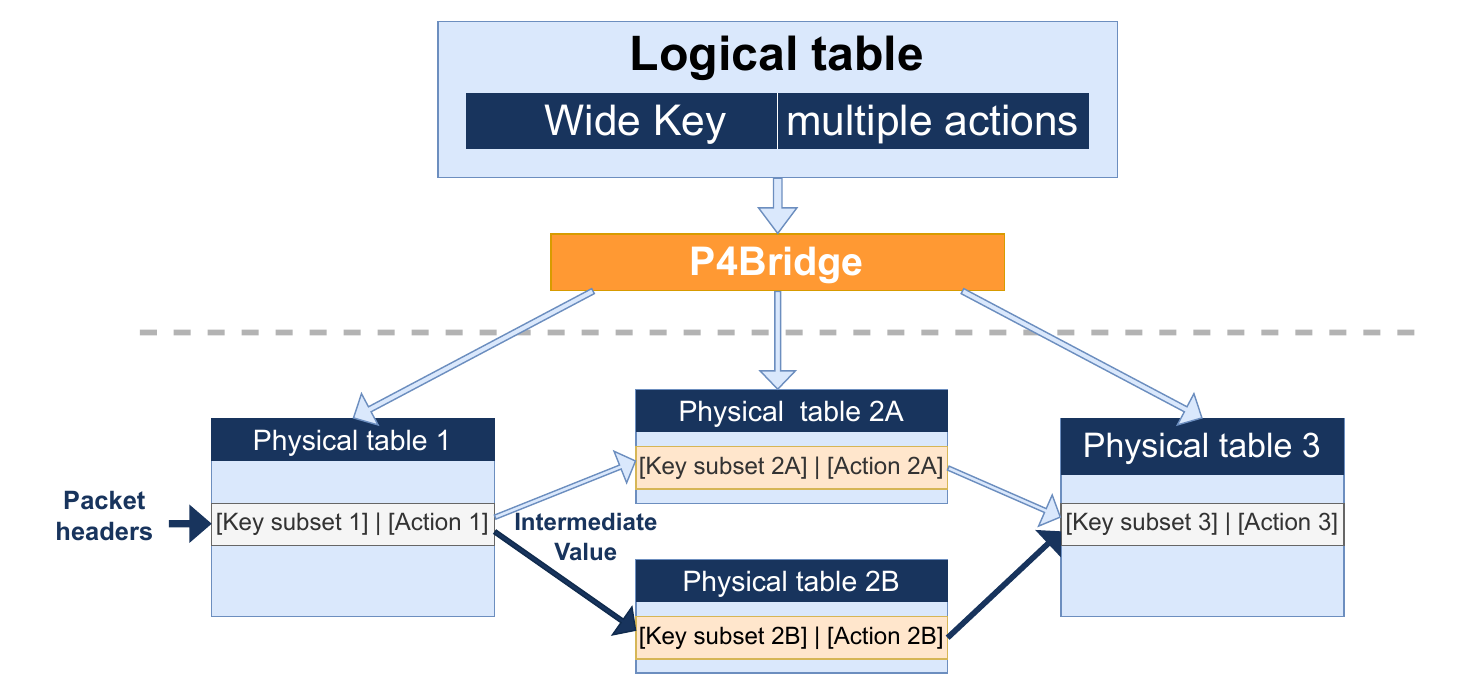} 
    \vspace{-0.15in}
    \caption{An example of table mapping via P4Bridge.}
    \vspace{-0.15in}
    \label{fig:p4bridge_abstraction}
\end{figure}

P4Bridge presents the gateway datapath as a single logical pipeline. Each
logical table can carry a wide key beyond hardware constraints, as shown in
Figure~\ref{fig:p4bridge_abstraction}, and P4Bridge maps each logical table
onto a set of hardware-resident physical tables, placing the corresponding
match keys and actions into the appropriate stages and installing the concrete
physical entries on the target hardware. Crucially, P4Bridge must preserve
logical-table semantics despite underlying table splitting, coalescing, and
remapping across ASICs and DPUs. P4Bridge also exposes consistent
configuration options for header handling (e.g., DSCP/TTL) and a unified
interface for counters and statistics. By hiding hardware-specific details
(e.g., parser differences and how tables are laid out or looked up), P4Bridge
lets control-plane developers program against a stable API while allowing
hardware engineers to optimize each backend transparently.

\subsection{Table Coalescing}
DPUs offer very limited TCAM, making hardware-efficient longest-prefix matching (LPM) impractical; even the Tofino ASIC's TCAM is not large enough to support all LPM operations \cite{pan2021sailfish, wang2019algorithmic}. Instead, when implementing prefix-based policies or routing rules using DPUs and Tofino SRAM, prefix-based matching is typically realized via hash-indexed exact-match lookups in SRAM/DRAM. A direct LPM-to-hash realization requires probing multiple candidate prefixes sequentially, and each probe may further consult an overflow structure to resolve collisions—multiplying memory accesses and control decisions~\cite{hanna2009progressive}. This overhead is directly manifested in the DPU datapath as additional pipeline stages and control logic, increasing per-packet processing cost and pressure. To mitigate this, we coalesce hash tables to reduce the number of probes and per-packet dispatch overhead. 

Our key observation is that, in cloud gateways, prefix overlaps are rare in
practice. \rev{We define two prefixes as overlapping when one prefix covers another
more-specific prefix. Prefixes that are non-overlapping under this definition
cannot produce competing LPM candidates, allowing P4Bridge to coalesce their
otherwise dependent hash lookups into a single table. In our production
configurations, such overlap is low, affecting only about 9.33\% of IPv4
prefixes and 6.63\% of IPv6 prefixes.}

Exploiting this property, the control plane performs conflict detection before
installing rules: if a set of prefixes can be guaranteed non-overlapping under
our matching semantics, we consolidate their entries into a single hash table
(and a shared overflow structure if needed). We further provide configuration
guidelines that encourage non-overlapping policies and maximize coalescing
opportunities. In the common case without multiple matches, coalescing
substantially reduces both lookup complexity and memory footprint on DPUs.

\subsection{Version-based Update} Table updates are particularly challenging in heterogeneous hardware environments. Forwarding rules span the Tofino pipeline, the DPU pipeline, and off-chip memory, each exposing different state representations and update primitives (e.g., fast-path tables updated via table-management interfaces vs. software-resident state updated through general-purpose memory operations under software-defined consistency) \cite{kim2023exoplane,michel2021programmable}. Synchronizing these views consistently is non-trivial, especially given the mismatch in update timescales. 

To this end, we keep mutable state in DRAM and run synchronization on the DPU's ARM cores, leveraging standard control-plane concurrency primitives. We compute a per-tenant service ID (SvcID) at the beginning of the Tofino pipeline and carry it as metadata to the DPU pipeline. Each service object is also tagged with a version number that reflects the latest configuration. The fast path only performs a lightweight version validation; upon detecting an expired version, it redirects the packet to the slow path for recomputation (Path 3 in Fig.~\ref{fig:pipeline}). After recomputation completes, the slow path installs the updated table entries into the fast path with the latest version. Upon tenants' reconfiguration, the controller increments the version in the Tofino-side table. Version-based update eliminates the need for tightly synchronized updates across components, while adding only a single version comparison on the fast path.

%% file: content/deployment.tex
\begin{figure*}[!ht]
	\centering
    \begin{minipage}[t]{0.25\textwidth}
        \includegraphics[width=1\textwidth]{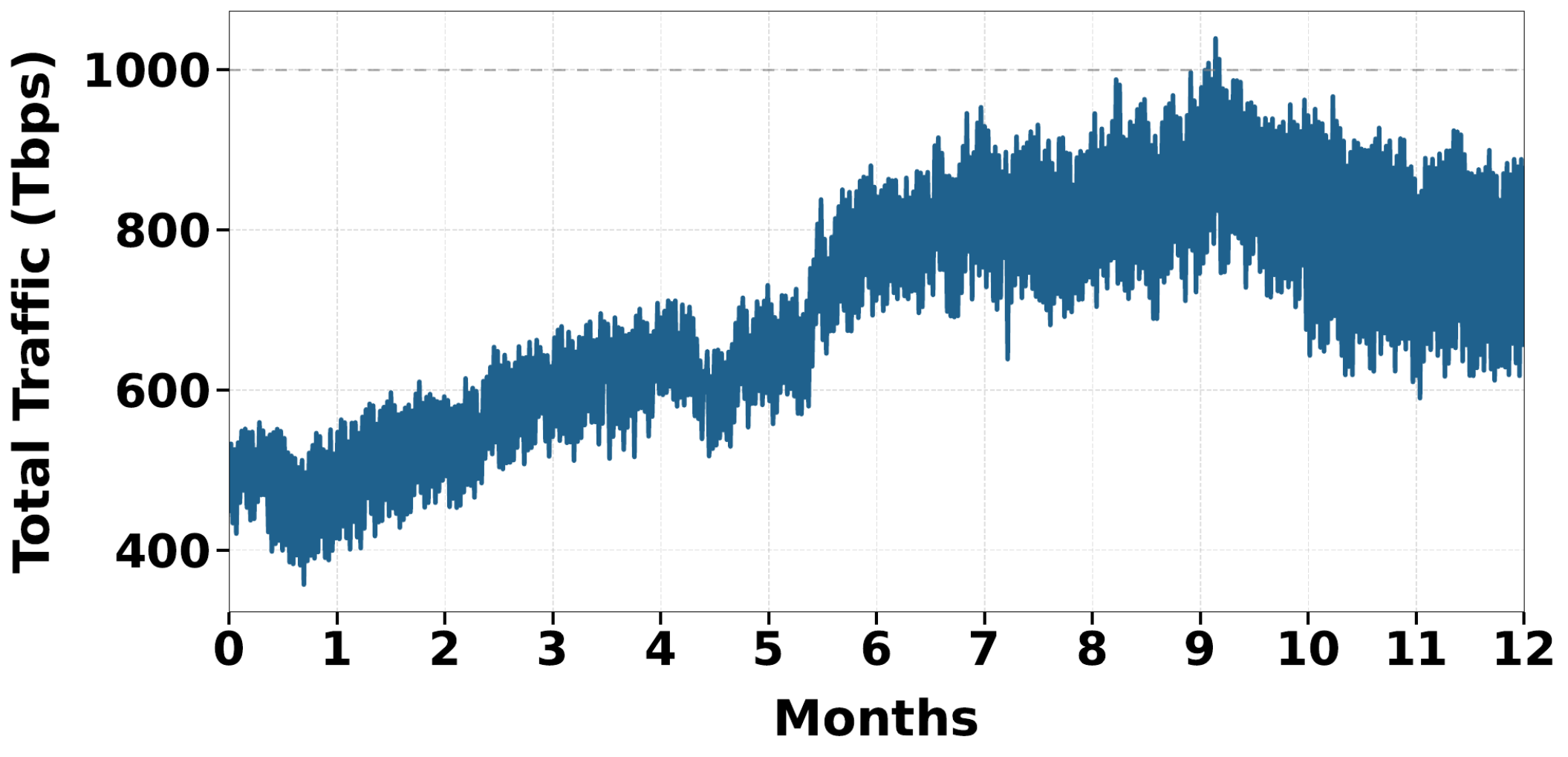}
        \vspace{-0.3in}
        \caption{Year-long production traffic.}
        \vspace{-0.1in}
        \label{exp:overall}
    \end{minipage}%
    \begin{minipage}[t]{0.25\textwidth}
        \includegraphics[width=1\textwidth]{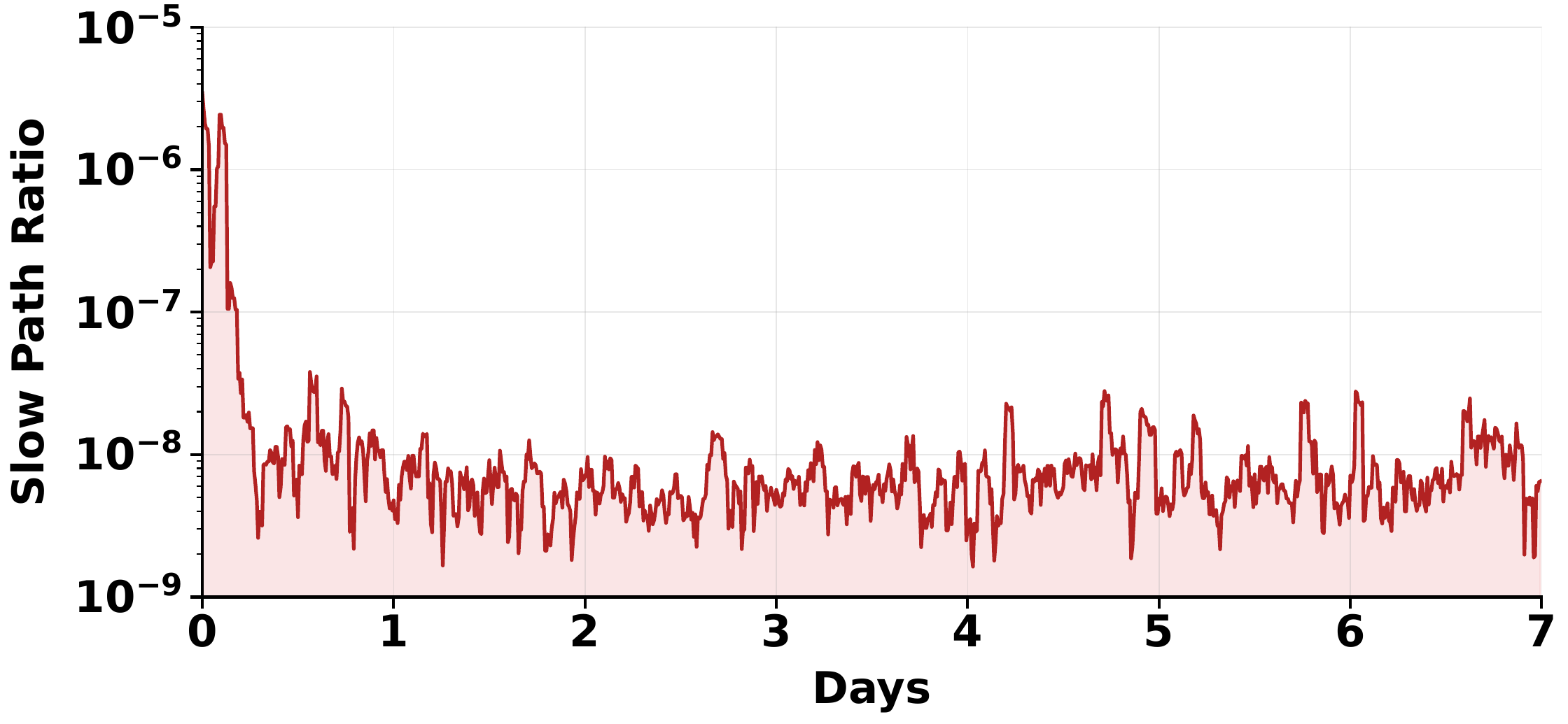}
        \vspace{-0.3in}
        \caption{Slow path traffic ratio.}
        \vspace{-0.1in}
        \label{exp:slow}
    \end{minipage}%
    \begin{minipage}[t]{0.25\textwidth}{
		\includegraphics[width=1\textwidth]{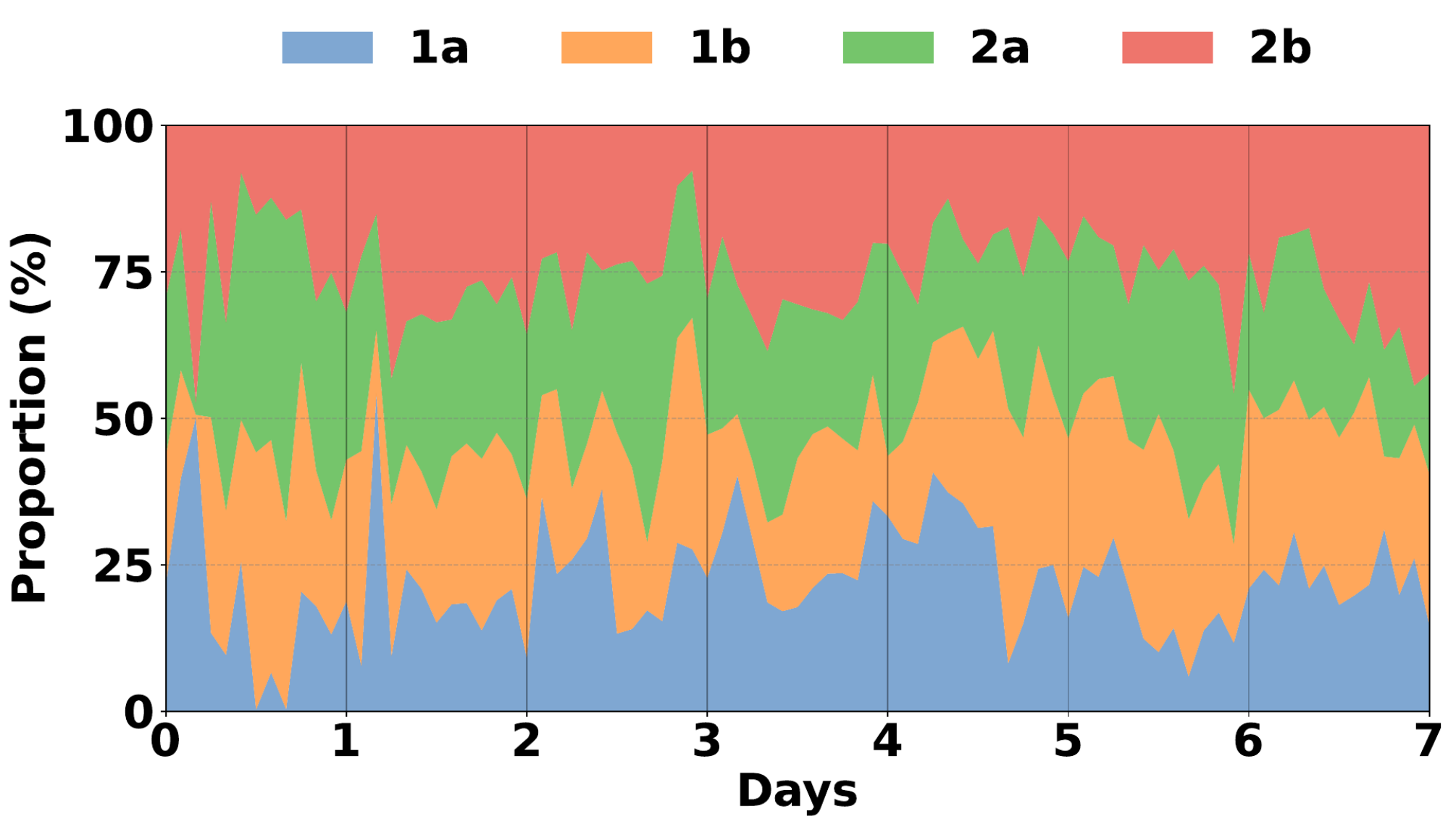}
        \vspace{-0.3in}
        \caption{Intra-DPU traffic distribution.}
        \vspace{-0.1in}
        \label{exp:hash_port}
	}\end{minipage}%
    \begin{minipage}[t]{0.25\textwidth}{
		\includegraphics[width=1\textwidth]{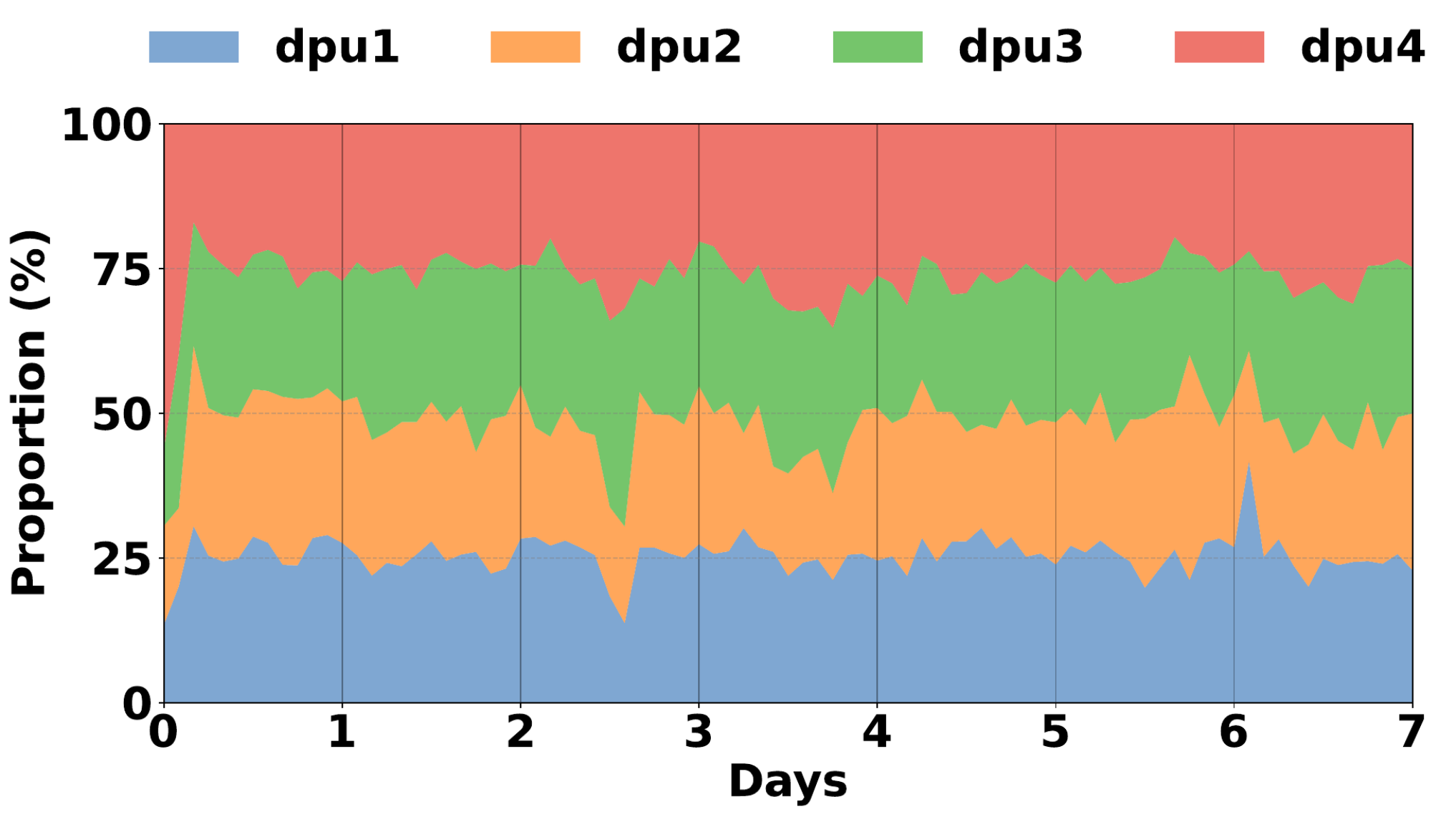}
        \vspace{-0.3in}
        \caption{Inter-DPU traffic distribution.}
        \vspace{-0.1in}
        \label{exp:hash}
	}\end{minipage}
\end{figure*}

\begin{figure*}[t]
  \centering
  \begin{minipage}[t]{0.32\textwidth}
    \centering
    \includegraphics[width=\linewidth]{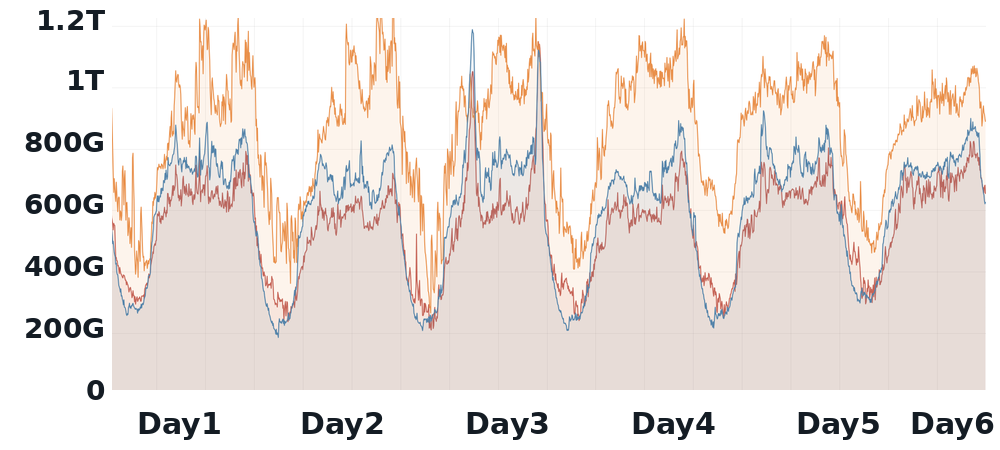}
    \vspace{-1.0em}
    \centerline{\rev{\small (a) Throughput.}}
  \end{minipage}
  \hfill
  \begin{minipage}[t]{0.32\textwidth}
    \centering
    \includegraphics[width=\linewidth]{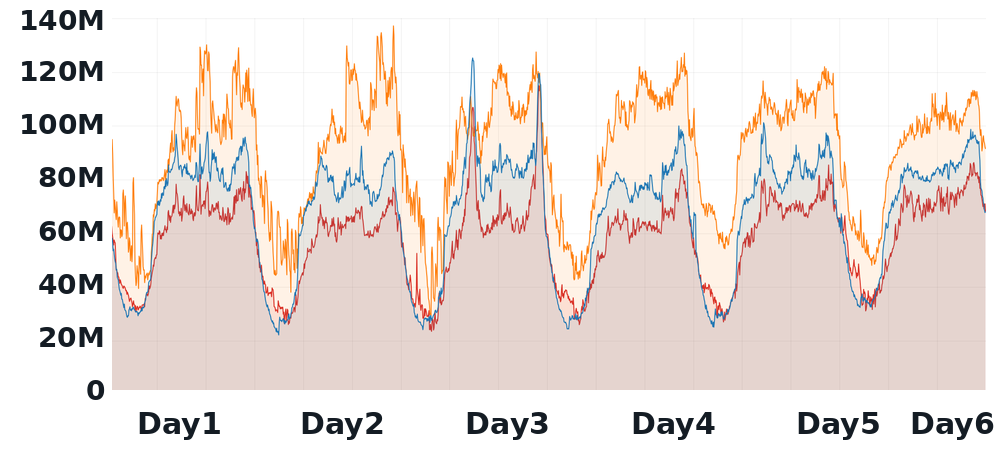}
    \vspace{-1.0em}
    \centerline{\rev{\small (b) Packet rate.}}
  \end{minipage}
  \hfill
  \begin{minipage}[t]{0.32\textwidth}
    \centering
    \includegraphics[width=\linewidth]{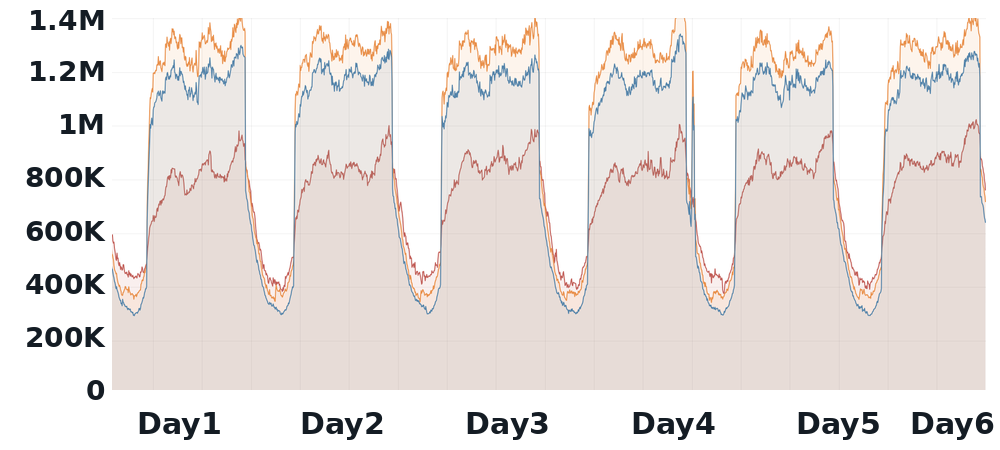}
    \vspace{-1.0em}
    \centerline{\rev{\small (c) Active flows.}}
  \end{minipage}

  \vspace{-0.1em}
  \caption{\rev{Workload statistics from one representative production cluster.}}
  \label{fig:production-workload-traces}
  \vspace{-1.0em}
\end{figure*}

\begin{table}[t]
\centering
\small
\caption{\rev{Production traffic distribution across \ourname{}'s hardware fast paths.}}
\vspace{-0.1in}
\label{tab:path_distribution}

\begin{tabular}{ll}
\toprule
Path & Relative share \\
\midrule
Switch-only path & $\sim$17\% \\
Hybrid fast path & $\sim$83\% \\
\bottomrule
\end{tabular}
\vspace{-0.2in}

\end{table}
\begin{figure*}[!ht]
    \centering
    \includegraphics[height=1.3em]{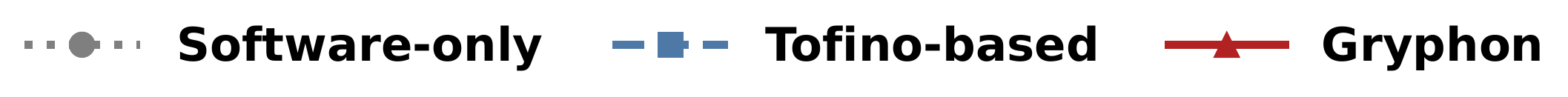} \\
    \vspace{-0.05in}
    \subfigure[Packet Loss Rate]{
	\begin{minipage}[t]{0.25\textwidth}{
		  \includegraphics[width=1\textwidth]{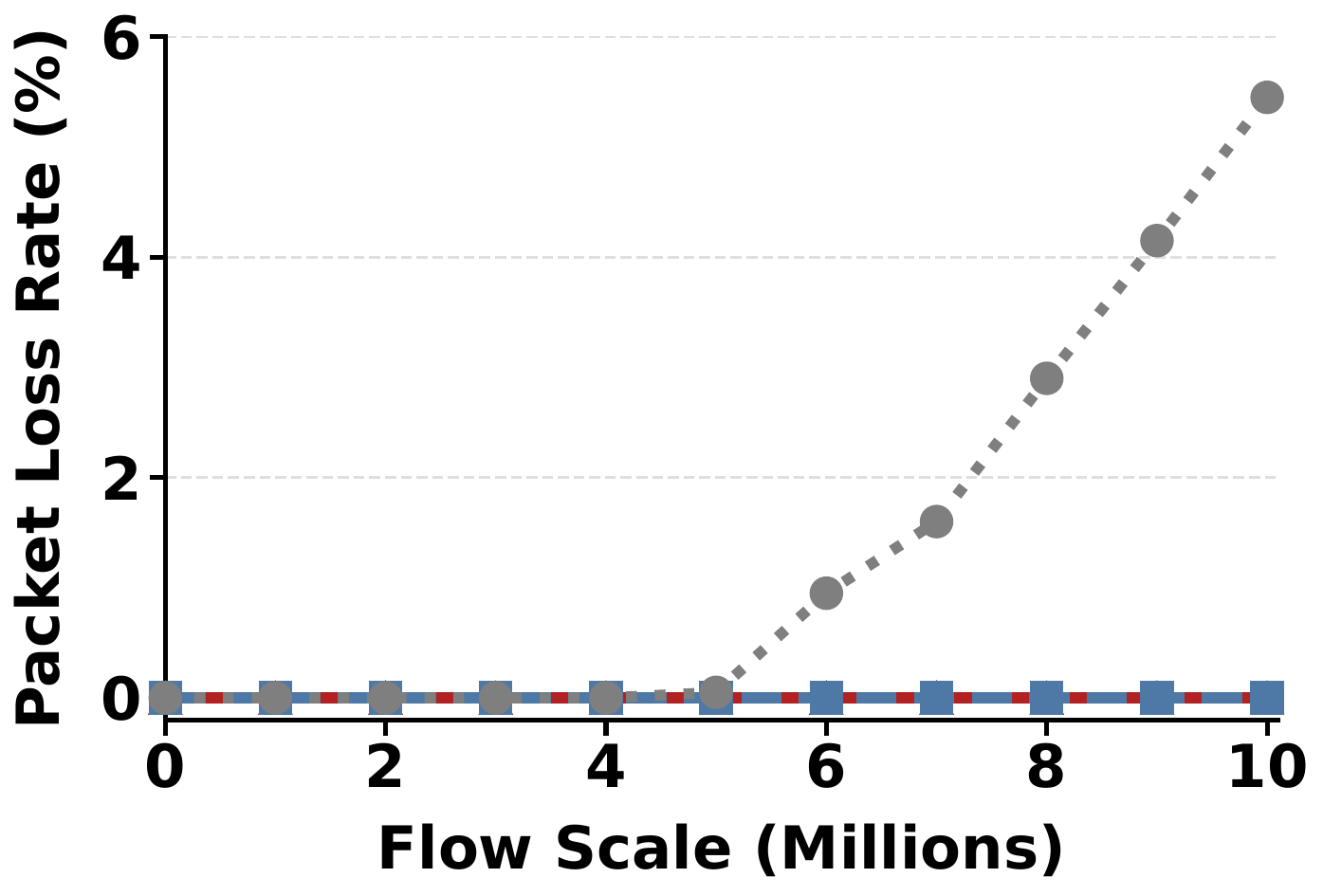}
        \label{exp:packet-loss}
        \vspace{-0.1in}
		}
	\end{minipage}}%
	\subfigure[Average Latency]{
	\begin{minipage}[t]{0.25\textwidth}{
		  \includegraphics[width=1\textwidth]{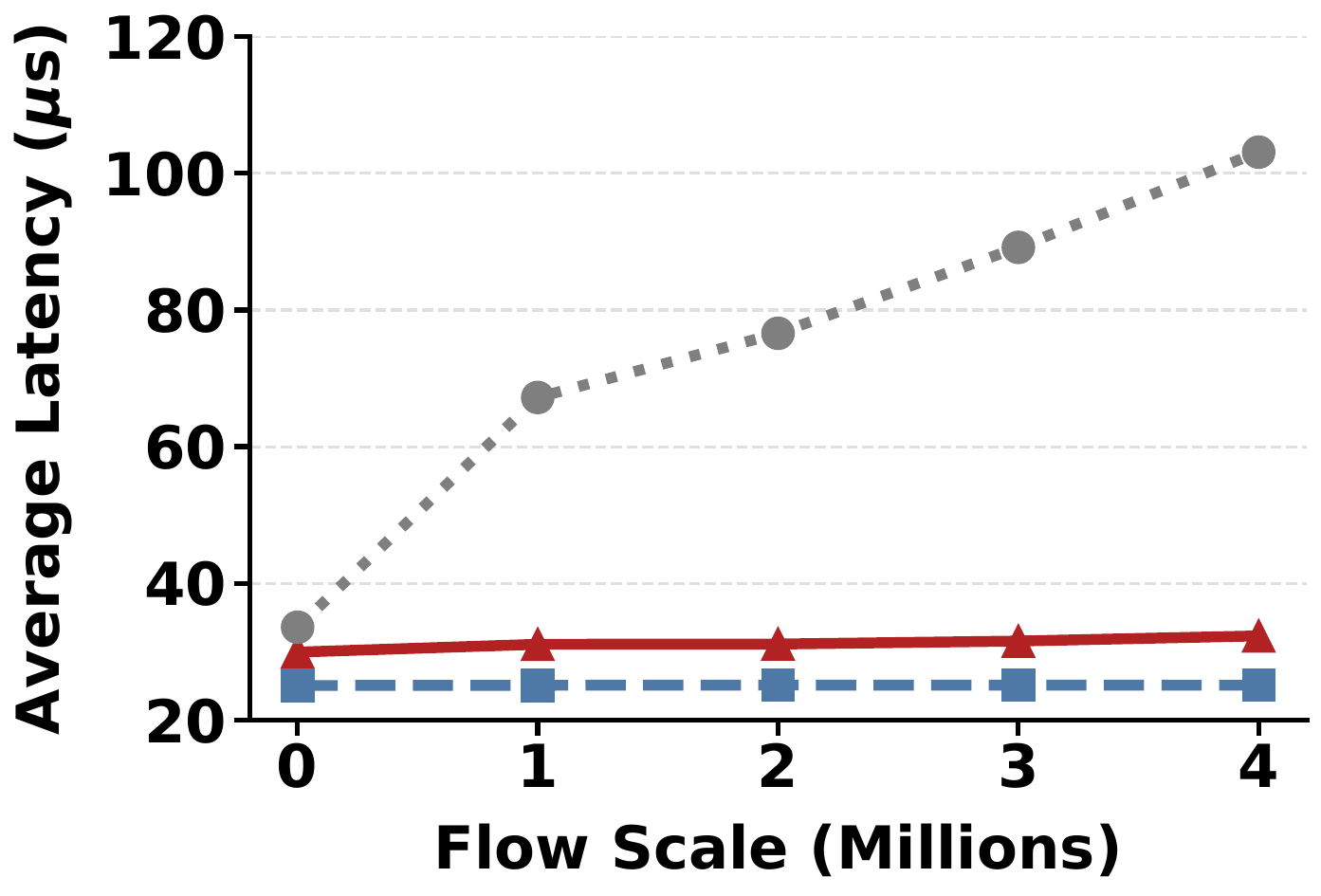}
          \label{exp:average-latency}
          \vspace{-0.1in}
		}
	\end{minipage}}%
	\subfigure[P99 Latency]{
	\begin{minipage}[t]{0.25\textwidth}{
		  \includegraphics[width=1\textwidth]{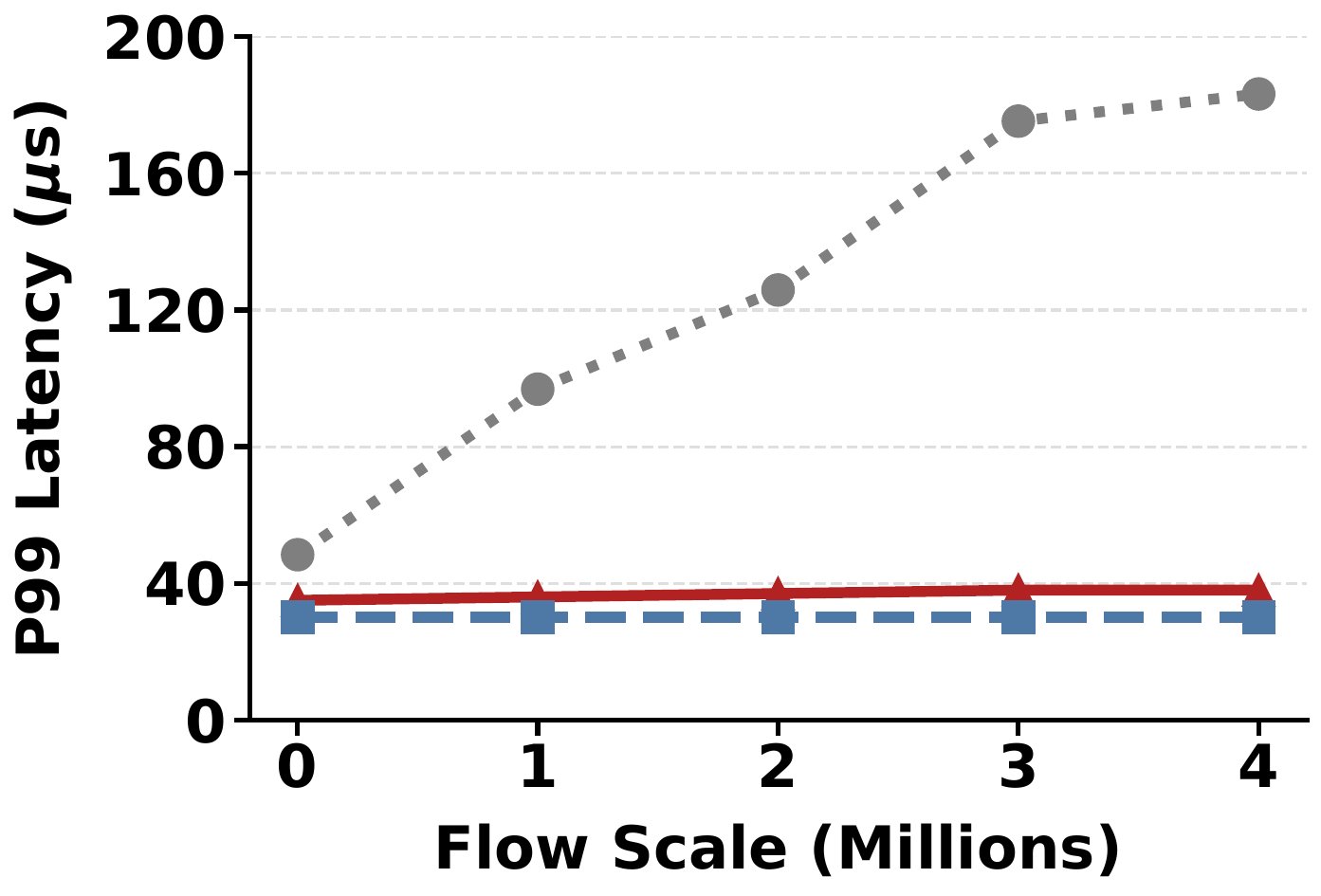}
          \label{exp:p99-latency}
          \vspace{-0.1in}
		}
	\end{minipage}}%
    \vspace{-0.1in}
	\caption{Performance comparison of three cloud gateway versions}
    \vspace{-0.1in}
\end{figure*}

\begin{figure*}[!ht]
	\centering
    \includegraphics[height=1.3em]{figure_exp/legend.pdf} \\
    \vspace{-0.05in}
    \subfigure[1K]{
	\begin{minipage}[t]{0.25\textwidth}{
		  \includegraphics[width=1\textwidth]{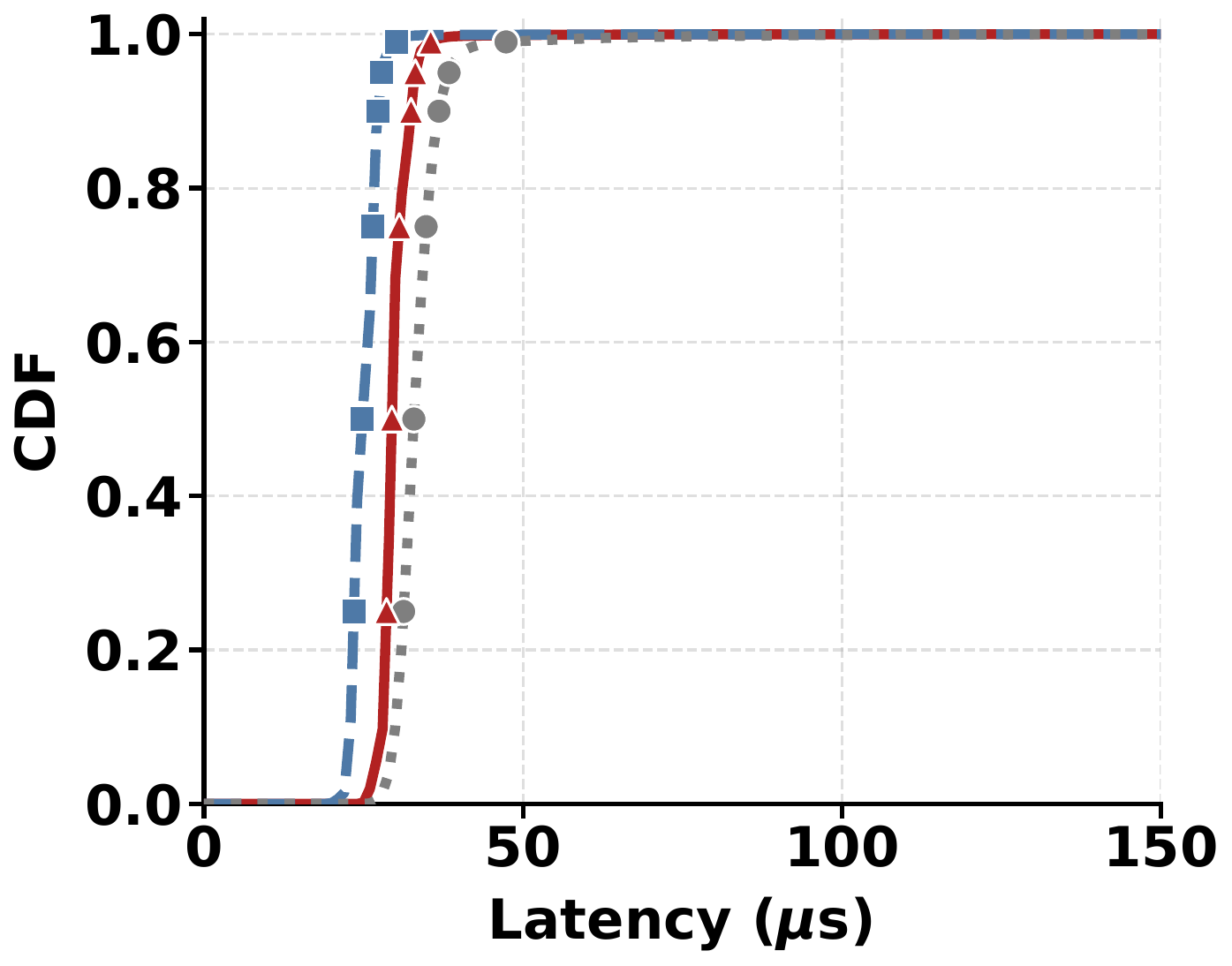}
          \vspace{-0.1in}
		}
	\end{minipage}}%
	\subfigure[1M]{
	\begin{minipage}[t]{0.25\textwidth}{
		  \includegraphics[width=1\textwidth]{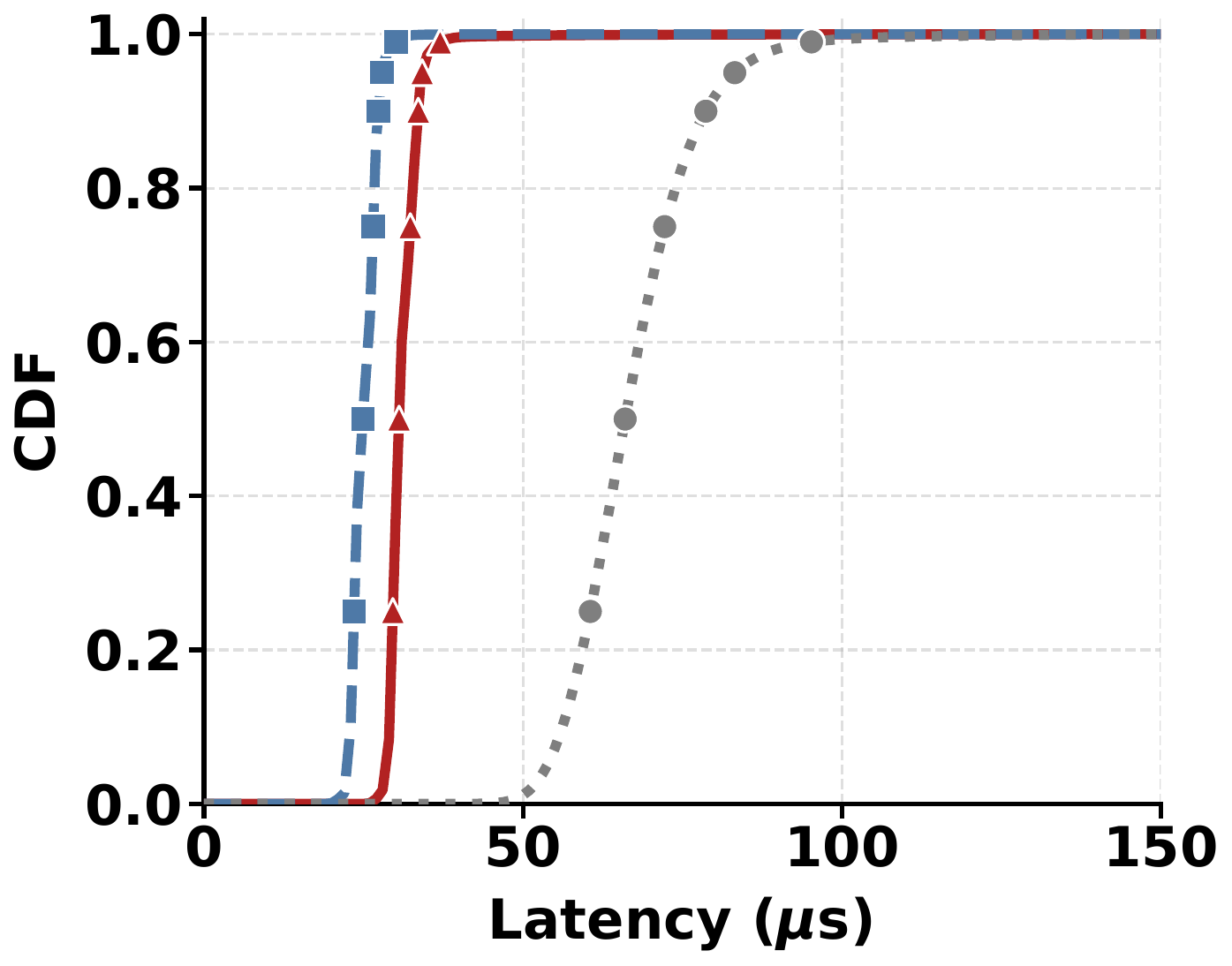}
          \vspace{-0.1in}
		}
	\end{minipage}}%
	\subfigure[4M]{
	\begin{minipage}[t]{0.25\textwidth}{
		  \includegraphics[width=1\textwidth]{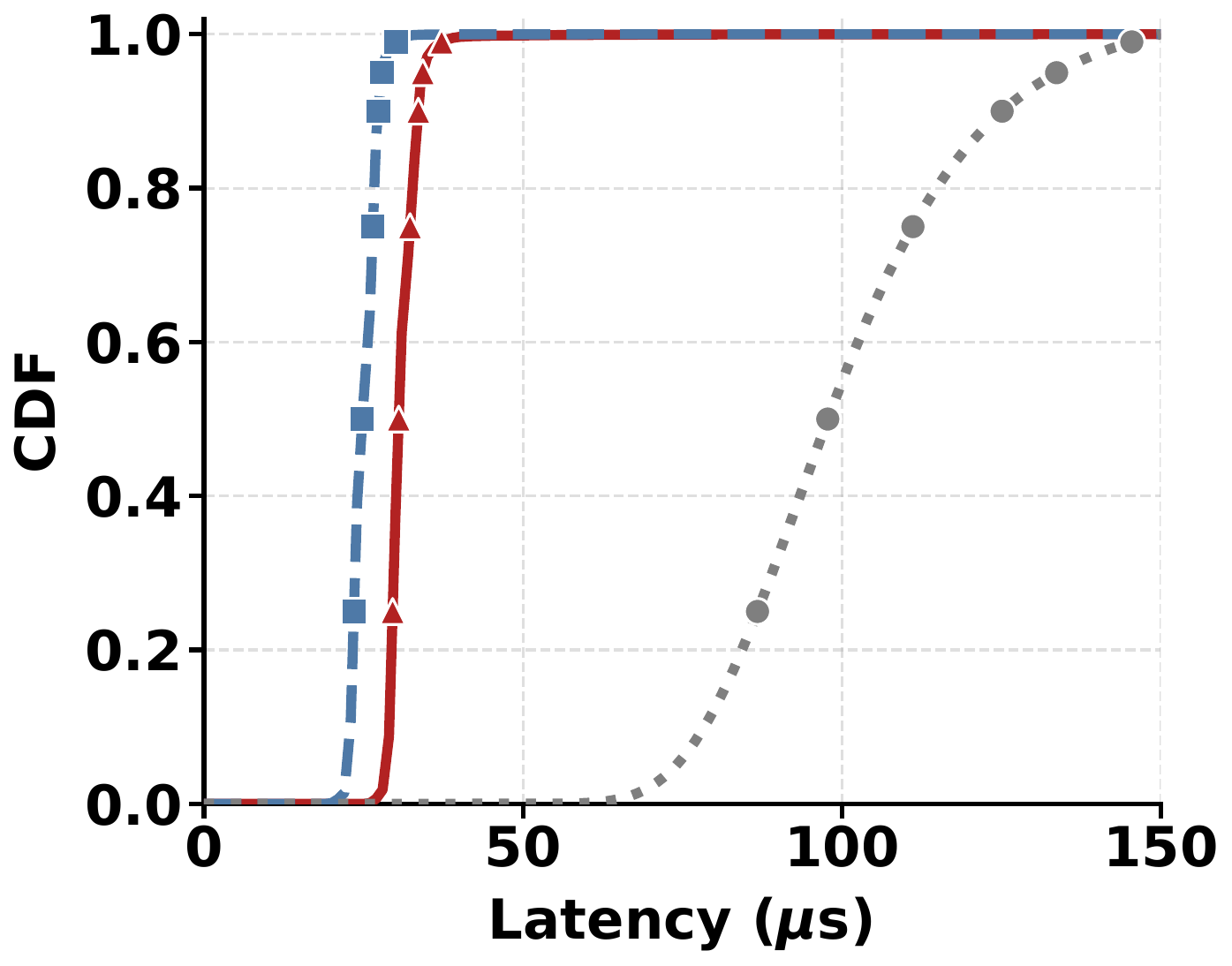}
          \vspace{-0.1in}
		}
	\end{minipage}}%
    \vspace{-0.1in}
	\caption{Latency distribution with different concurrent flow scales}
    \label{exp:cdf}
    \vspace{-0.1in}
\end{figure*}

\section{Deployment}




\ourname{} has been deployed in production as part of ByteDance’s cloud gateway infrastructure \cite{volcengine}, supporting both north--south and east--west traffic.
It is currently deployed across hundreds of gateway nodes within multiple clusters in five availability zones (AZs), with each node featuring a folded Tofino ASIC coupled with four Pensando DPUs.



The system has operated continuously in production for over one year, handling a significant portion of production traffic and sustaining multi-terabit-per-second aggregate throughput under high-concurrency workloads.
The deployment continues to expand, gradually replacing previous-generation gateways that cannot meet current performance and scalability demands, with its footprint continuously growing as it integrates into an increasing number of cloud services and workloads. 

%% file: content/implementation+.tex
\section{Implementation}
We implemented \ourname{} on an in-house gateway platform~\cite{edgecore_wedge100bf_qsg_r01}. The system is equipped with an Intel Tofino 6.4 Tbps ASIC featuring 64 $\times$ 100 Gbps ports. We integrated four AMD Pensando Elba DPUs~\cite{pensando2022}, each featuring a 16-core ARM Cortex-A72 SoC for localized, complex processing. To bridge the DPU and the switching ASIC, we utilized breakout Active Optical Cables (AOCs) to split each 200 Gbps DPU port into two 100 Gbps links. This "sandwich" configuration ensures full-duplex bandwidth and aligns the DPU interconnect with the gateway's 1.6 Tbps line-rate throughput requirements.

%% file: content/implementation_new.tex
\section{Evaluation}


\subsection{Real-world Traffic Characterization}

We present real-world traffic measurements collected from our production deployment to demonstrate the scale and operational characteristics of \ourname{}. 

\textbf{Year-long production traffic.}
Figure~\ref{exp:overall} illustrates the aggregate traffic throughput over a one-year period.
The data reveals a steady upward trend, peaking at over 1 Pbps.
The sheer magnitude of this traffic underscores the critical need for a gateway architecture to deliver massive throughput beyond the capabilities of traditional software-only solutions.

\begin{figure*}[t]
  \centering
  
  \includegraphics[width=0.35\linewidth]{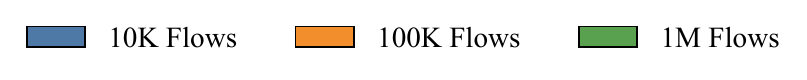}
  \vspace{0.2em} 
  
  \centerline{
    \begin{minipage}[t]{0.34\textwidth}
      \centering
      \includegraphics[width=\linewidth]{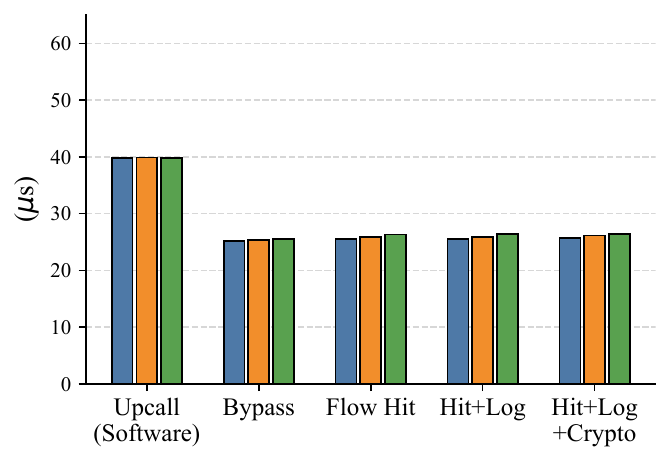}
      \vspace{-1.2em}
      \centerline{\small (a) Average latency.}
    \end{minipage}
    \hfill
    \begin{minipage}[t]{0.34\textwidth}
      \centering
      \includegraphics[width=\linewidth]{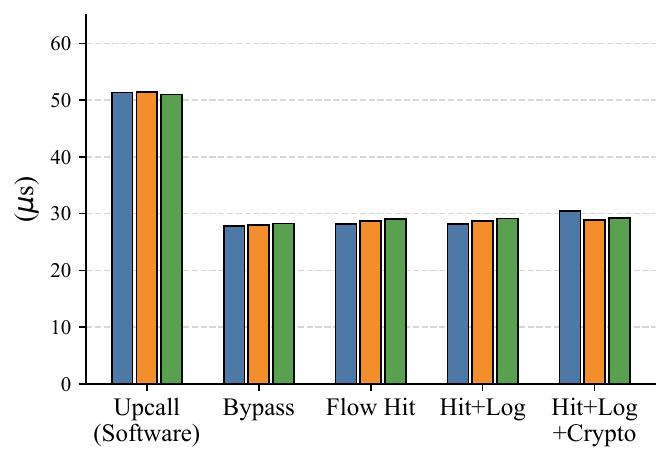}
      \vspace{-1.2em}
      \centerline{\small (b) P99 latency.}
    \end{minipage}
    \hfill
    \begin{minipage}[t]{0.34\textwidth}
      \centering
      \includegraphics[width=\linewidth]{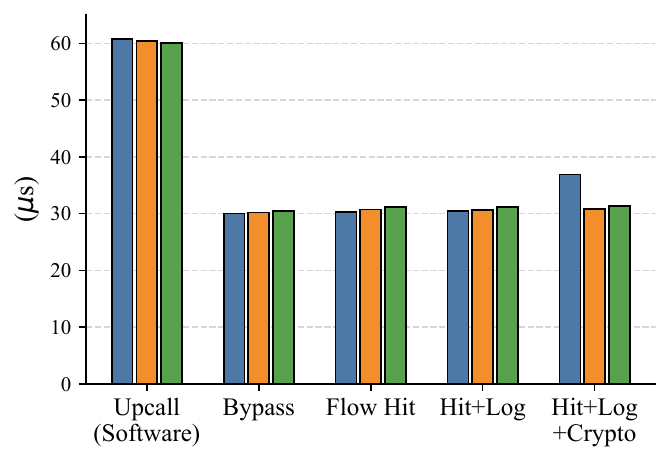}
      \vspace{-1.2em}
      \centerline{\small (c) P99.9  latency.}
    \end{minipage}
  }

  \vspace{-0.1em}
  \caption{\rev{DPU forwarding latency under different pipeline configurations and flow scales. \textit{Flow Hit} is the hardware fast-path baseline; \textit{Hit+Log} additionally writes per-flow counters; \textit{Hit+Log+Crypto} further enables inline encryption/decryption.}}
  \label{fig:dpu_latency}
  \vspace{-1.5em}
\end{figure*}

\rev{
\textbf{Six-day production workload.}
Figure~\ref{fig:production-workload-traces} presents production workload
traces collected over a six-day period. The traces exhibit clear
diurnal patterns while remaining at consistently high load: cluster
throughput reaches the Tbps scale, packet rate reaches hundreds of
Mpps, and flow pressure remains persistently high. This
confirms that production gateways must simultaneously handle high
bandwidth, high packet rate, and large forwarding-state pressure,
rather than optimizing for only one dimension.
}

\rev{
\textbf{Production path distribution.}
To better understand \ourname{}'s production behavior, we analyze the
traffic distribution across the three paths defined in Section~\ref{subsec:high_level_architecture}:
the Switch-only path, the Hybrid fast path, and the Software slow path.
We find that slow-path traffic is mainly triggered by transient events such
as bursts of new-flow arrivals, after which it quickly returns to a low
steady level. As illustrated in Figure~\ref{exp:slow}, the Software slow
path consistently accounts for a very small fraction of traffic, staying
below 0.001\% for the vast majority of the time. This observation is
consistent with our hierarchical co-offloading design, which keeps the
common cases in hardware and punts only rare cases to software. Moreover,
the Software slow path in our current deployment processes packet headers
only, without inspecting payloads, which bounds per-packet work and reduces
bandwidth and compute contention, making it less likely to become a
bottleneck even when overall traffic is high.
}

\rev{
We further classify a
production traffic snapshot across the two hardware fast paths. As shown in
Table~\ref{tab:path_distribution}, the resulting split is approximately
17:83: only about 17\% of traffic follows the Switch-only path, whereas the
remaining 83\% uses the Hybrid fast path and therefore benefits from
DPU-resident state capacity or inline processing. This workload composition
shows that pure ASIC forwarding is important but not dominant in production
cloud gateways; most traffic still relies on the DPU tier for large tables,
or advanced gateway functions.
}

\textbf{DPU load balance.}
As illustrated in Figure~\ref{exp:hash_port}, intra-DPU traffic distribution is susceptible to skew in our production environment. 
Such imbalances are often inevitable due to our per-flow pinning policy, which ensures that all packets in a flow follow an identical path to prevent reordering.
To mitigate this, we employ a dual-mode steering strategy that reactively rebalances traffic across the gateway complex.
Figure \ref{exp:hash} demonstrates the effectiveness of this approach: 
the stacked area chart shows that the load is evenly distributed across all four DPUs over a one-week period, with each DPU consistently handling approximately 25\% of the aggregate traffic (additional data for other nodes are provided in Appendix~\ref{app:load}). 


\begin{figure*}[!ht]
	\centering
    \includegraphics[height=1.15em]{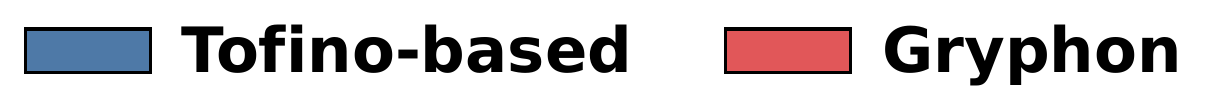} \\
    \vspace{-0.05in}
    \subfigure[Pipeline 0]{
	\begin{minipage}[t]{0.24\textwidth}{
		  \includegraphics[width=1\textwidth]{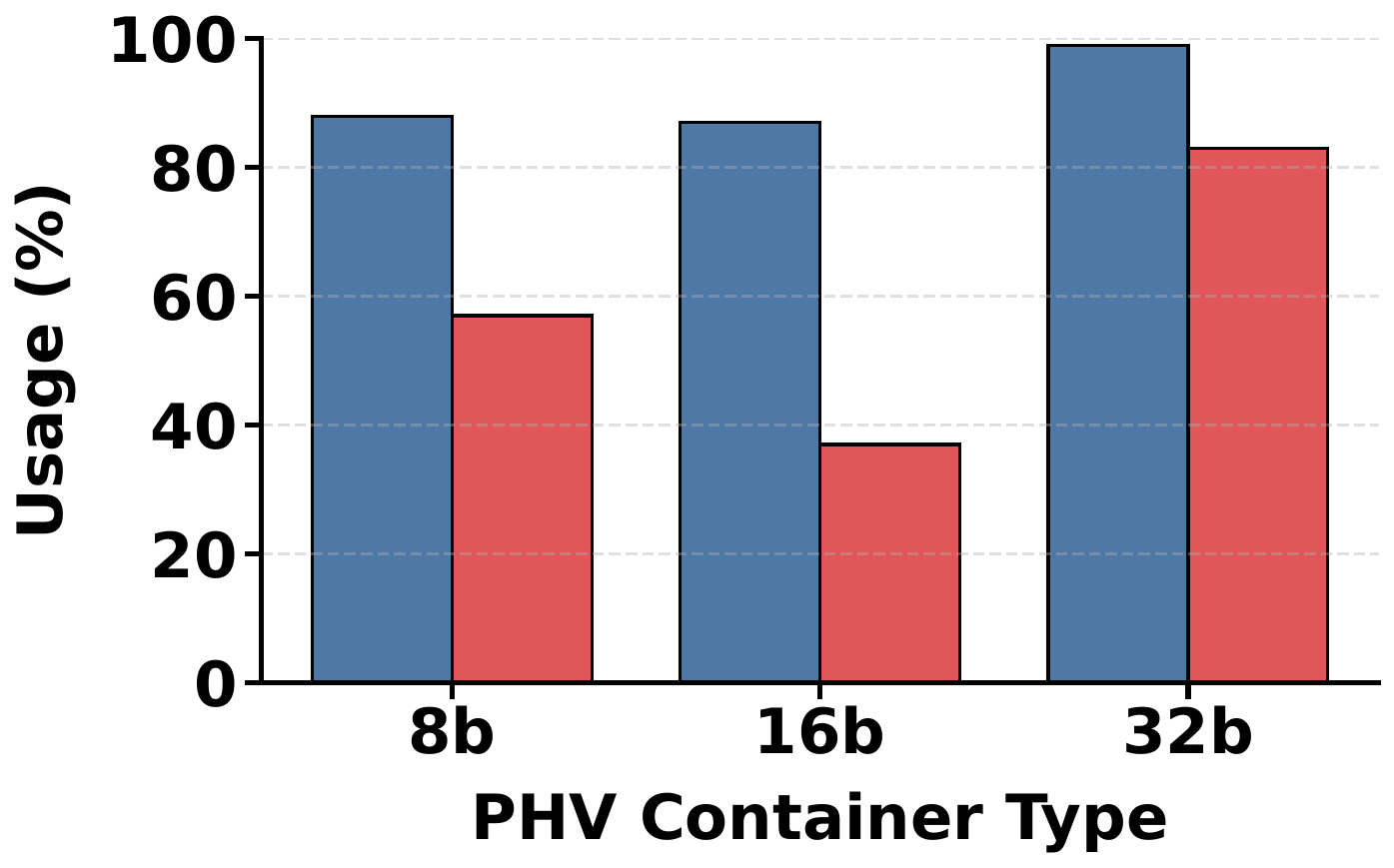}
          \vspace{-0.1in}
		}
	\end{minipage}}%
	\subfigure[Pipeline 1]{
	\begin{minipage}[t]{0.24\textwidth}{
		  \includegraphics[width=1\textwidth]{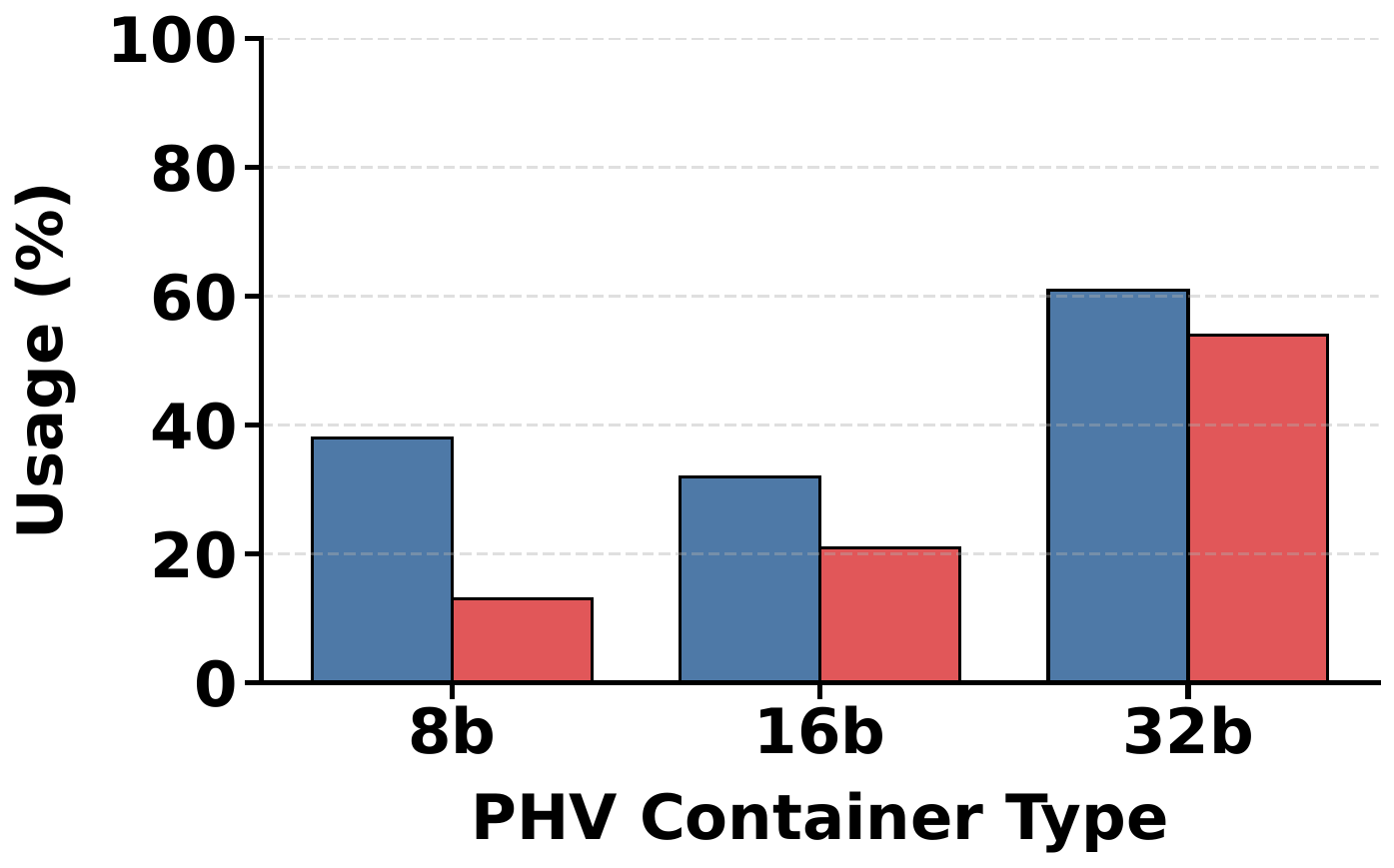}
          \vspace{-0.1in}
		}
	\end{minipage}}%
	\subfigure[Pipeline 2]{
	\begin{minipage}[t]{0.24\textwidth}{
		  \includegraphics[width=1\textwidth]{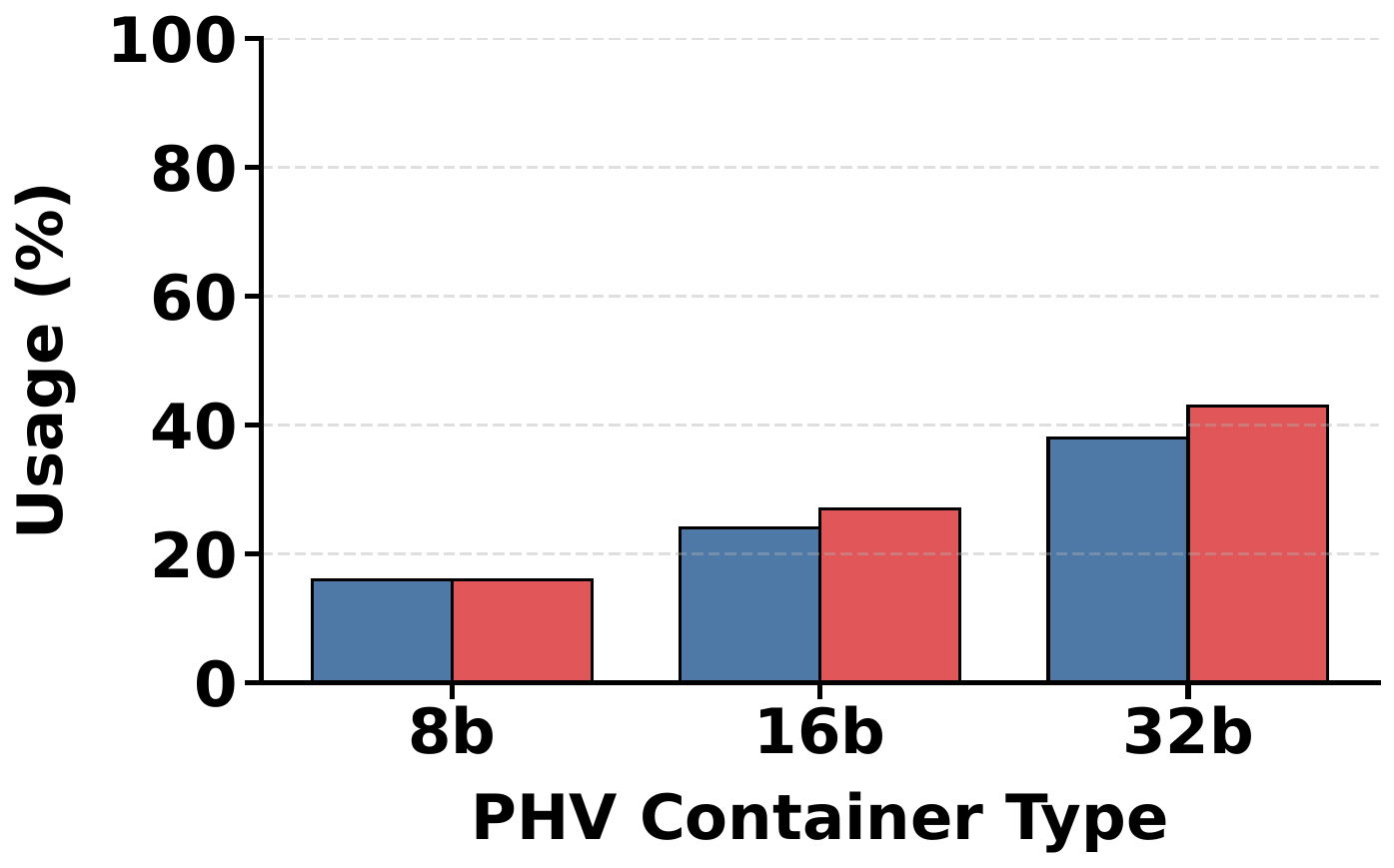}
          \vspace{-0.1in}
		}
	\end{minipage}}%
    \subfigure[Pipeline 3]{
	\begin{minipage}[t]{0.24\textwidth}{
		  \includegraphics[width=1\textwidth]{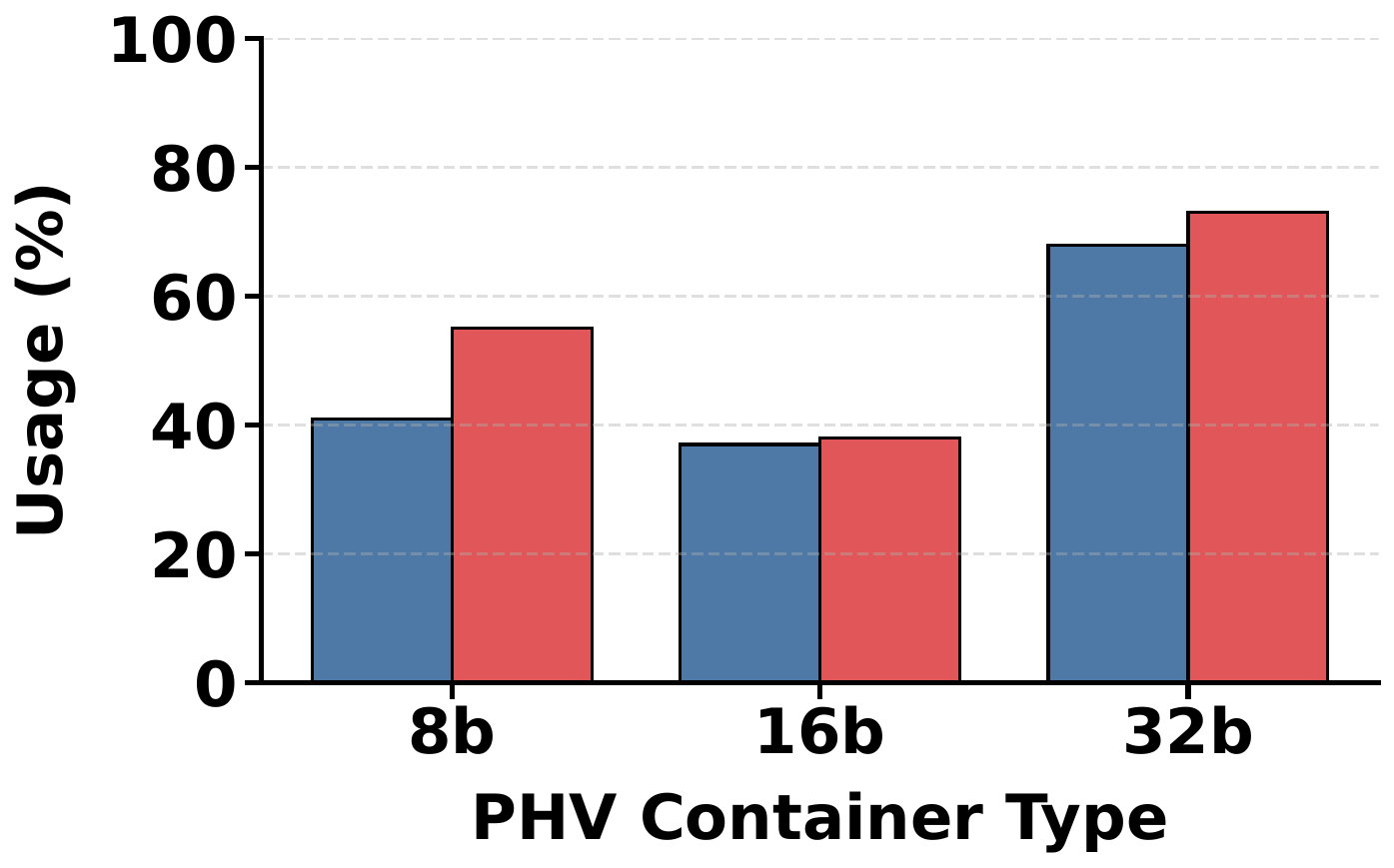}
          \vspace{-0.1in}
		}
	\end{minipage}}%
    \vspace{-0.1in}
	\caption{PHV usage across pipelines (the worst case).}
    \label{exp:phv}
\end{figure*}

\begin{figure*}[!ht]
	\centering
    \begin{minipage}[t]{0.5\textwidth}
        \includegraphics[width=1\textwidth]{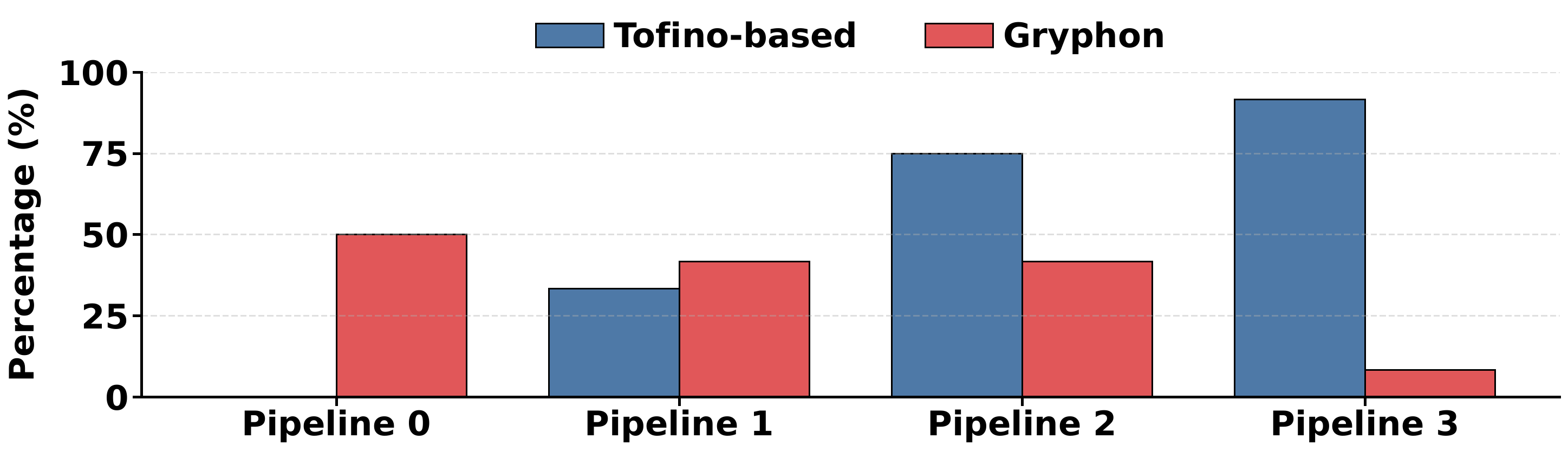}
        \vspace{-0.3in}
        \caption{SRAM saturation percentage across pipelines.}
        \vspace{-0.1in}
        \label{exp:satu}
    \end{minipage}%
    \begin{minipage}[t]{0.5\textwidth}{
		\includegraphics[width=1\textwidth]{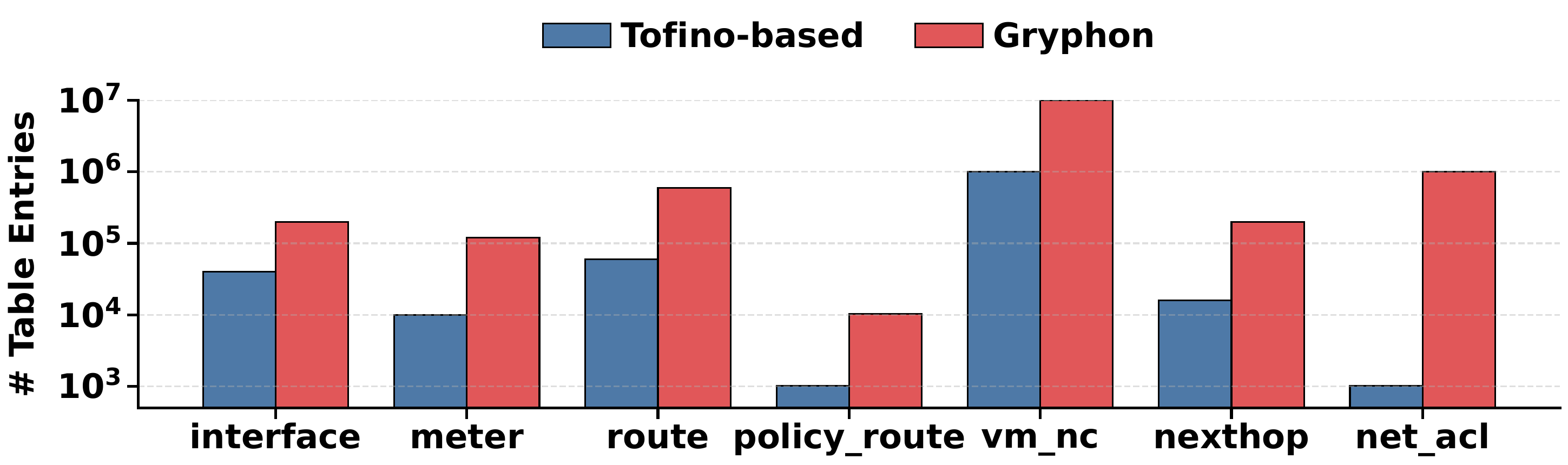}
        \vspace{-0.3in}
        \caption{Table capacity of different versions.}
        \vspace{-0.1in}
        \label{exp:table}
	}\end{minipage}
\end{figure*}

\begin{table*}[t]
\centering
\caption{Comparison of efficiency metrics (normalized to best-in-class = 1.00$\times$).}
\vspace{-0.1in}
\label{tab:efficiency-comparison-normalized}
\small
\setlength{\tabcolsep}{3pt}
\begin{tabular}{lccccccccc}
\toprule
\multirow{2.5}{*}{\textbf{Cloud Gateway}} & 
\multirow{2.5}{*}{\textbf{\begin{tabular}[c]{@{}c@{}}Table\\ capacity\end{tabular}}} & 
\multirow{2.5}{*}{\textbf{\begin{tabular}[c]{@{}c@{}}Throughput\\ (Gbps)\end{tabular}}} & 
\multicolumn{2}{c}{\textbf{Per device}} & 
\multicolumn{4}{c}{\textbf{Efficiency (Normalized, 1.00$\times$ is Best)}} & 
\multirow{2.5}{*}{\textbf{Space}} \\ 

\cmidrule(lr){4-5} \cmidrule(lr){6-9}

 &  &  & 
\textbf{Power} & \textbf{Cost} & 
\textbf{\begin{tabular}[c]{@{}c@{}}Power/\\ Tbps\end{tabular}} & 
\textbf{\begin{tabular}[c]{@{}c@{}}Power/\\ Entry\end{tabular}} & 
\textbf{\begin{tabular}[c]{@{}c@{}}Cost/\\ Tbps\end{tabular}} & 
\textbf{\begin{tabular}[c]{@{}c@{}}Cost/\\ Entry\end{tabular}} & 
 \\ 
\midrule


Software-only & $>10\times$ & $\sim$200 & 1.11$\times$ & 1.00$\times$ & 
\textbf{8.89$\times$} & 1.00$\times$ & 
\textbf{6.67$\times$} & 1.00$\times$ & 
2U \\ \addlinespace

Tofino-based & 1$\times$ & $\sim$1600 & \textbf{1.00$\times$} & 1.20$\times$ & 
1.00$\times$ & \textbf{>9.00$\times$} & 
1.00$\times$ & \textbf{>12.00$\times$} & 
2U \\ \addlinespace

\ourname{} & $\mathbf{>10}\times$ & $\mathbf{\sim1600}$ & 2.08$\times$ & 3.64$\times$ & 
2.08$\times$ & 1.87$\times$ & 
3.03$\times$ & $\sim$3.64$\times$ & 
2U \\ 

\bottomrule
\end{tabular}%
\end{table*}

\subsection{Performance}

To evaluate \ourname{}'s operational boundaries, we conduct tests in a real production environment. 
By sweeping the number of active flow entries from 1K to 10M, we measure packet loss rate and end-to-end latency to assess how each design manages massive flow states.

\textbf{Packet loss.}
Figure \ref{exp:packet-loss} illustrates the packet loss rate as the flow scale expands. 
Both the Tofino-based gateway and \ourname{} maintain zero packet loss throughout the sweep, proving their ability to handle 10M concurrent flows stably.
In contrast, the software-only gateway suffers a significant performance collapse once the flow scale exceeds 5M. 
Beyond this point, the overhead of managing high-density flow tables leads to packet drops exceeding 5\% at peak scale, highlighting the limitations of software-based forwarding when flow states exceed the capacity of local CPU caches.

\textbf{End-to-end latency.}
We evaluate latency within the stable operating region (up to 4M flows) where all candidates exhibit zero packet loss, which allows for a fair comparison of the steady-state forwarding performance.
The results (Figures \ref{exp:average-latency}--\ref{exp:p99-latency}) show that \ourname{} substantially reduces both average and P99 latency compared to the software-only gateway.
At a flow scale of 4M, \ourname{} reduces the average latency by 68.7\% and the P99 latency by 79.3\%. 
In addition, \ourname{} adds only a modest ~8$\mu$s delay relative to the Tofino-only design. 
The near-flat latency profile in Figure \ref{exp:cdf} confirms that \ourname{} provides the deterministic performance needed for latency-sensitive services, effectively avoiding the jitter and timeouts typically seen in software solutions under heavy flow loads.

\textbf{Per-Function DPU Overhead.}
We isolate the per-packet overhead introduced by each inline
function on the DPU by measuring average, P99, and
P99.9 latencies across five pipeline configurations
and three flow scales (10K, 100K, 1M), as shown in
Figure~\ref{fig:dpu_latency}.  

\textit{Flow Hit}---in which matching flows are forwarded entirely
within the DPU hardware pipeline---serves as the baseline for this
breakdown.

\textit{Software upcall overhead.}
When a packet falls back to the software upcall path, the average
latency rises to ${\sim}40\,\mu$s and the P99.9 latency exceeds
$60\,\mu$s, representing ${\sim}14\,\mu$s ($1.5\times$) and
${\sim}30\,\mu$s ($2\times$) overhead over \textit{Flow Hit},
respectively.  This gap is likely influenced by overheads in hardware--software transitions.

\textit{Flow-logging overhead.}
Enabling per-flow logging (\textit{Hit+Log}) appends a small
number of counter writes to the existing hardware pipeline.
Compared to \textit{Flow Hit}, this adds less than $0.1\,\mu$s on
average and less than $0.2\,\mu$s at P99.9 across all flow
scales, indicating that flow logging is effectively free at the
hardware forwarding level.

\textit{Inline-encryption overhead.}
Further enabling encryption and decryption
(\textit{Hit+Log+Crypto}) leverages the DPU's dedicated on-chip
crypto engines.  At 100K and 1M flows, the additional overhead
beyond \textit{Hit+Log} remains within $0.2$--$0.3\,\mu$s on
average and under $0.3\,\mu$s at P99.9.  At 10K flows, we
observe a higher tail-latency spike, while the average overhead
remains modest at $+0.1$--$0.2\,\mu$s.  This suggests that
inline crypto is lightweight in the common case, with occasional
tail events at smaller flow scales.
Across hardware-offloaded configurations, average latency grows
by less than $1\,\mu$s as the flow count scales from 10K to 1M,
demonstrating stable performance under increasing flow pressure.


\subsection{Hardware Efficiency and Scalability}


In this section, we evaluate the hardware efficiency of \ourname{} by comparing its resource usage and table capacity against the Tofino-based gateway.

\textbf{PHV utilization.}
Figure \ref{exp:phv} compares PHV utilization for the Tofino-based cloud gateway and \ourname{}. 
The results show that 32-bit PHV constrains the scalability of the Tofino-based gateway: in the worst case, pipeline 0 nearly exhausts its 32-bit PHV capacity. 
In contrast, \ourname{} maintains a safe scaling margin across all three PHV widths (8/16/32-bit) and pipelines; in every pipeline, utilization for each width remains below 85\%, enabling further scaling.

\textbf{SRAM utilization.}
We define stages with SRAM usage > 95\% as
saturated, and plot the percentage of such stages for each pipeline
in Figure~\ref{exp:satu}. For the Tofino-based gateway, over
90\% of stages (11 out of 12) in pipeline 3 are saturated, whereas for
\ourname{} the fraction of saturated stages stays below 60\% in every
pipeline. The relatively higher utilization observed in a few stages is
caused by compiler placement and resource packing after adding
metadata and steering logic, rather than by a new capacity bottleneck.
Overall, \ourname{}'s SRAM utilization remains healthy, providing a
substantial margin for growth and reducing the risk of resource scarcity.

\textbf{Table capacity.}
Figure \ref{exp:table} summarizes the maximum table capacities for the Tofino-based cloud gateway and \ourname{}. 
Leveraging DPUs' massive memory, \ourname{} supports substantially more entries --- up to 10M \texttt{gw\_vm\_nc} entries versus 1M on the Tofino-based gateway. 
\ourname{} also offers at least 10$\times$ more entries for \texttt{gw\_interface}, \texttt{gw\_meter}, \texttt{gw\_route}, and \texttt{gw\_nexthop}, and 1000$\times$ for \texttt{gw\_net\_acl}. 
In addition, \ourname{} supports extra tables such as \texttt{gw\_nexthop\_affinity}, which are infeasible on the Tofino-based gateway.

\textbf{Head-to-head comparison: trade-off.} Our analysis reveals a shifting bottleneck across generations:
(1) In the first stage, software gateways offered massive table capacity ($>10\times$) but were severely limited by throughput ($\sim$200 Gbps).
(2) In the second stage, Tofino-based gateways resolved the throughput deficit ($\sim$1600 Gbps) but hit a bottleneck in table capacity.
Simplistic scale-out incurs prohibitive overheads in efficiency: Scaling software to match Tofino's throughput requires roughly 8$\times$ the rack space, 8.89$\times$ power consumption and 6.67x procurement cost. While scaling switching ASICs to match software capacity results in unacceptable Power/Entry ($>9.00\times$), Cost/Entry ($>12.00\times$), and Space/Entry.
Notably, \ourname{} resolves this dilemma through high-density integration. By consolidating four DPUs into the 2U chassis, \ourname{} achieves $\sim$1600 Gbps throughput and $>10\times$ table scale simultaneously.
This density prevents the prohibitive Space/Entry costs, allowing \ourname{} to serve not only massive public clouds but also space-constrained edge clouds~\cite{pan2024luoshen}, where physical rack space is a scarce and rigid resource.

\rev{Notably, \ourname{} is not designed to outperform a Tofino-only gateway in pure
line-rate forwarding. Instead, it preserves the folded-pipeline line rate
of $\sim$1.6 Tbps while providing $>10\times$ table capacity, substantially
better power and cost per entry than scaling Tofino-only gateways, and a
broader set of deployable gateway functions. Compared with our Tofino-only
gateway, \ourname{} enables seven additional functions, including
fragmented-packet offload and encryption. 
Thus, cloud gateways need DPUs primarily for flexible processing and large
state capacity, rather than for marginal raw-throughput gains over an
already line-rate ASIC datapath.}

%% file: content/relatedwork.tex
\section{Related Work}
\vspace{-0.3em}
Cloud vendors have explored different network virtualization architectures. Azure offloads host networking to SmartNICs~\cite{AzureSmartNIC, DBLP:conf/nsdi/FirestonePMCDAA18} to reduce CPU overhead and improve performance. Google's Andromeda uses Hoverboard programming
model, which uses gateways for the long tail of low bandwidth flows~\cite{DBLP:conf/nsdi/DaltonSAAGFRZRD18}. Recent work further explores disaggregated accelerator pools to absorb spillover network-function load beyond what a single host-attached accelerator can provide~\cite{bansal2023disaggregating}. 

For centralized cloud gateway, the industry has witnessed a clear evolution in data plane technologies. Early deployments, such as those by Alibaba Cloud, relied on software-based gateways running on x86 server clusters to handle north-south and east-west traffic~\cite{pan2021sailfish,pan2024luoshen}. 
However, due to CPU bottlenecks, these systems transitioned to programmable switching ASICs (e.g., Tofino) to achieve high throughput and cost efficiency~\cite{pan2021sailfish}. 
Sailfish~\cite{pan2021sailfish} represents this state-of-the-art switching ASIC-based approach, employing hardware-software co-design to mitigate the chip's limited on-chip memory. Luoshen~\cite{pan2024luoshen} further advances this paradigm by building a hyper-converged programmable gateway for multi-service edge clouds. However, as cloud scale continues to explode, the strict capacity bounds of on-chip resources eventually limit these solutions~\cite{lu2025albatross}.

To address this, other architectures explore FPGA-based acceleration. For example, Albatross~\cite{lu2025albatross} leverages FPGAs to implement packet-level load balancing (PLB) with containerized flexibility. Although solutions like Albatross offer larger table capacities than switching ASICs, FPGA development involves significantly longer development cycles compared to the P4-based workflow used in \ourname{}. Besides, its limited per-device throughput (800Gbps) potentially becomes a bottleneck for bandwidth-hungry workloads in the petabit era. \rev{Similarly, Tiara~\cite{zeng2022tiara} proposes a hardware-accelerated architecture for stateful layer-4 load balancing that combines a high-throughput programmable switch for traffic steering with FPGA/HBM-based processing for scalable per-flow state management. In contrast, Gryphon targets hyperscale multi-tenant cloud gateways rather than a specific stateful L4 load-balancing service: it supports a broader set of cloud-gateway functions, including tenant routing, ACLs, metering, VM--NC mapping, encapsulation/decapsulation, and optional inline functions such as flow logging and encryption.}

By introducing DPUs into the pipeline, \ourname{} overcomes the resource constraints of ASIC-only gateways like Sailfish and Luoshen, while avoiding the development complexity and higher costs associated with FPGA-based designs like Albatross. This positions \ourname{} as a viable and efficient candidate for next-generation cloud infrastructures.

%% file: content/experience.tex
\section{Lessons Learned}

\textbf{Mitigating hash-induced DPU hotspots.} Initially, we assumed that hash-based steering~\cite{google:maglev} would be sufficient to balance load across DPUs; however, deployment traffic skew~\cite{Hedera-NSDI10, DBLP:conf/sigcomm/AlizadehEDVCFLMPYV14} repeatedly concentrated heavy-hitter flows on the same DPU. To address this, we implemented two \rev{complementary} modes for steering traffic
across DPUs. \rev{By default, packets are steered to DPUs by hashing flow metadata to select
a DPU-facing Tofino port. Under hash skew, \ourname{} switches to one-to-one mapping, where selected
traffic groups are pinned to specific DPU-facing ports.} This port redirection facilitates flow rebalancing, helping the
inter-DPU load reach a balanced state. We found the two modes complementary
in operations, with hashing in the common case and deterministic pinning
activated only during skew episodes.

\textbf{DPU-aware offloading.}
Although DPUs significantly extend the storage capacity and programmability of switching ASICs, our experience shows that naively migrating dataplane functions to the DPU can significantly degrade forwarding performance if architectural placement and implementation are not carefully designed. \textbf{Case 1 (Prefix matching on DPU):} 
Initially, we followed the common design intuition~\cite{DBLP:journals/corr/abs-2504-03653, kfoury2024smartnic} that the DPU’s massive off-chip DRAM could be effectively utilized to host large prefix tables and implement LPM. However, LPM requires multiple dependent probes per packet, turning each lookup into iterative off-chip DDR accesses. Although lookups can be served from fast on-chip SRAM, the added DDR accesses quickly dominated processing latency, significantly degrading performance. Therefore, we moved prefix matching back to the switching ASICs and implemented it using Tofino's ALPM in on-chip SRAM~\cite{pan2021sailfish}. \textbf{Case 2 (metering on DPU):} We implemented a token-bucket meter on the DPU for rate limiting. \rev{Our initial meter implementation used a compact fused arithmetic expression for token refresh, but this caused the compiler to generate expensive wide-operand arithmetic, resulting in more than 60 instructions. By rewriting the computation in a compiler-aware form and keeping intermediates narrow, we reduced the meter to about 40 instructions.}

\rev{
\textbf{Scaling processing-heavy offloads.}
A lesson from \ourname{} is that the DPU tier should be treated not only as
expanded memory for large tables, but also as an elastic processing tier.
While our current deployment is primarily driven by table-capacity pressure,
many gateway features can become compute-bound as operators add richer
inline functions. Compute-heavy operations such as encryption and
compression/decompression are therefore
better placed on DPU accelerators or programmable DPU pipelines than on
switching ASICs. More importantly, \ourname{}'s multi-DPU
steering makes such processing capacity scalable: traffic can be spread
across stronger DPUs or across more DPU cards per node without changing
the logical control-plane abstraction. This design makes \ourname{} applicable
not only to memory-bound gateways, but also to processing-bound
network functions.
}

\textbf{P4Bridge fits best in gateway designs with 5-tuple-based fast/slow path separation.}
In our deployment, we find that P4Bridge aligns well with pipelines that use 5-tuple matching and implement fast/slow path selection. In such cases, due to the constraints on the maximum match key size of a single P4 table, a single control-plane logic table is typically decomposed into multiple dataplane tables, resulting in a natural 1:n mapping that reflects the pipeline’s structural constraints. By contrast, in designs which use ASIC-only forwarding and non-5-tuple-based matching—logic tables tend to follow a simpler 1:1 mapping with dataplane tables. This simplicity may weaken P4Bridge’s value in practice. Although more aggressive table merging (i.e., $m$:1 mappings) improves hardware resource utilization, the resulting increase in development complexity may undermine the design goal of P4Bridge of accelerating development. As a result, we adopt the simpler 1:1 mapping in non-5-tuple-based cases, which maintains the clarity of the logical API model and avoids overcomplicating the P4Bridge's development.

\textbf{DPUs for faster iteration and debuggability than FPGA offload.}
We initially explored an FPGA-based offload~\cite{lu2025albatross} design for the same class of gateway functions.
While FPGAs offer strong programmability and can host large tables, we found the end-to-end engineering loop substantially slower in practice: changes often span RTL, drivers, host software, and simulation/emulation \cite{DBLP:journals/tcad/GoedersW17,DBLP:conf/fpga/AttiaB20}, and a single feature iteration can take on the order of half a year from design to validated deployment. Such long cycles make timely delivery difficult for us in a fast-moving production environment.
In contrast, vendor DPUs provide a more integrated hardware/software stack and mature tooling,
which improves iteration velocity and debuggability for production rollout~\cite{kfoury2024smartnic,li2024triton}.
It motivated our decision to prioritize DPU-based offloading for productization while keeping the control-plane-facing abstractions stable to accommodate future hardware changes.

\textbf{P4Bridge stabilizes control-plane interfaces during iterations.} Due to strict hardware constraints and performance objectives, the implementation of the DPU pipeline is often deeply coupled with physical resources. Architectural evolution——whether for refactoring, optimization, or introducing new features such as encryption and flow logging——necessitates reshaping the physical table layout (e.g., splitting or merging P4 tables to combine or decompose functionality during refactoring or when adding modules such as flow logging and encryption). Exposing these pipeline- and vendor-specific table APIs to control-plane developers creates brittle dependencies: bottom-up layout changes repeatedly force control-plane code changes. Our operational experience confirms that \ourname{} (P4Bridge) shields upper layers from this churn. Across multiple production iterations (e.g., P4 table adjustments and adding flow logging/encryption), the logical pipeline exposed to the control plane remained stable. Moreover, during our migration from BlueField~\cite{DBLP:journals/cn/LeeYKP26} to Pensando, P4Bridge successfully masked hardware- and vendor-specific differences across DPUs. Adapting the abstraction layer required substantially less engineering and collaborative effort than rewriting upper-layer control logic against the DPU's native interface.

\textbf{Future-proofing the dataplane across switching ASIC generations.}
We initially built \ourname{} on Intel Tofino, taking advantage of its native P4 match-action abstraction~\cite{DBLP:journals/ccr/BosshartDGIMRSTVVW14}. 
At hyperscale, however, hardware churn is the norm: as deployments expand and the ASIC ecosystem evolves, we must be able to retarget the dataplane to other switches (e.g., Broadcom Trident~5~\cite{broadcom_trident5_bcm78800}) without a ground-up rewrite. 
To this end, we implement a P4-like dataplane in NPL~\cite{broadcom_trident5_bcm78800}. Since NPL’s table-side expressiveness differs from Tofino’s per-entry action binding, we adopt a disciplined convention: each P4 action is encoded as an \emph{action id}, its parameters are exported as metadata fields, and a hardware FSL (Flexible Switch Logic) block—supporting predicate evaluation and arithmetic—dispatches and executes the corresponding behavior.
Based on our engineering experience with the Broadcom Trident 5, we have found that the P4-like NPL substantially reduces porting effort and shortens iteration cycles; 
combined with P4Bridge, it also keeps control-plane changes minimal when retargeting to different switching ASICs. Appendix~\ref{sec:appendix_p4like_npl} presents a concrete example.

\textbf{\ourname{}'s general applicability.} Initially, the community's consensus has often focused on scaling out ASIC-only clusters to meet growing demands~\cite{google:jupiter, pan2021sailfish,pan2024luoshen}. However, in our environment, petabit-scale aggregate traffic and explosive growth of forwarding rules push both software gateways and ASIC-only designs to their limits. From the perspective of small and medium-sized cloud operators, simpler designs may be “good enough” today—many deployments may not yet reach hyperscale table pressure or require a scaling architecture. But as workloads grow and multi-tenant isolation demands more state, the gap between required table capacity and what on-chip resources can provide will widen; moreover, scaling out ASIC-only gateways fragments state and complicates traffic steering and synchronization. We therefore believe the design space \ourname{} targets --- tight ASIC–DPU coupling for “capacity + performance” rather than a pure software/ASIC approach—is not a niche solution, but a direction that more operators will likely need sooner or later~\cite{lu2025albatross,kim2023exoplane,DBLP:journals/corr/abs-2504-03653}.

%% file: content/Conclusion.tex

\section{Conclusion}
\ourname{} is ByteDance's third-generation cloud gateway. DPU integration, together with the hierarchical co-offloading strategy, overcomes challenges and limitations of previous gateways, without compromising their core strengths. Built for the petabit era, \ourname{} scales to support massive traffic workloads while meeting emerging business needs. We are continuously optimizing this new architecture to deliver higher performance and meet emerging business demands. As we move forward with the system's continuous upgrades and deployments, we are committed to sharing our operational experiences and insights with the cloud networking community.

\section*{Acknowledgement} 
We would like to thank the anonymous reviewers and the shepherd for their constructive comments.
This work was supported by the National Key Research and Development Program of China under Grant No. 2024YFB2906602, and in part by the National Natural Science Foundation of China (NSFC) (No. 62372009). 
Generative AI tools were used to assist with language polishing and grammatical improvements.

%% file: content/appendix.tex
\section{A Concrete Example of our P4-like NPL}
\label{sec:appendix_p4like_npl}

In P4, a match-action table can directly bind multiple actions with distinct parameters. 
In NPL, we encode the action choice as an \emph{action id} (e.g., \texttt{action\_ctrl\_0}) and export all action parameters as table fields, then implement an action-dispatch function in the hardware logic unit (FSL) to realize the corresponding behavior.

\subsection{P4 (Tofino) Program Snippet}

\begin{lstlisting}[style=sigcommcode]
#define CONST_IP 0x0A0A0A0A

// action_hit
action fib_hit(inout switch_local_metadata_t local_md,
               switch_nexthop_t nexthop_index) {
  local_md.nexthop = nexthop_index;
  local_md.routed = true;
}

// action_1
action fib_myip_1(inout switch_local_metadata_t local_md,
                  switch_myip_type_t myip) {
  local_md.myip = myip;
  local_md.myflag = true;
}

// action_2
action fib_myip_2(inout switch_local_metadata_t local_md,
                  switch_myip_type_t myip) {
  local_md.myip = myip;
  local_md.myflag = false;
}

// action_3
action fib_myip_3(inout switch_local_metadata_t local_md,
                  bool myflag) {
  local_md.myip = CONST_IP;
  local_md.myflag = myflag;
}

// action_drop
action fib_drop(inout switch_local_metadata_t local_md) {
  local_md.routed = false;
  local_md.fib_drop = true;
}

table my_ipv4_host {
  key = {
    local_md.vrf : exact @name("vrf");
    local_md.lkp.ip_dst_addr[95:64] : exact @name("ip_dst_addr");
  }
  actions = {
    fib_hit;
    fib_myip_1;
    fib_myip_2;
    fib_myip_3;
    fib_drop;
  }
  const default_action = fib_drop(local_md);
  size = my_host_table_size;
}

control SwitchIngress(...) {
  my_ipv4_host.apply();
  ...
}
\end{lstlisting}

\subsection{P4-like NPL Implementation}

\begin{lstlisting}[style=sigcommcode]
P4 logical_table my_ipv4_host {
  table_type : hash;
  minsize    : my_host_table_size;
  maxsize    : my_host_table_size;

  keys {
    bit[FIELD_32_WD] ipv4_addr;
    bit[VRF_WD]      vrf;
  }

  // Export action choice + all potential parameters as fields.
  fields {
    bit[FIELD_4_WD]  action_ctrl_0;
    bit[FIELD_16_WD] nexthop_index;
    bit[FIELD_32_WD] myip;
    bit[FIELD_1_WD]  routed;
    bit[FIELD_1_WD]  myflag;
  }

  key_construct() {
    if (_LOOKUP0 == 1) {
      vrf       = ing_obj_bus.vrf;
      ipv4_addr = pkt_fwd_field_bus.ip_hdr_dip[31:0];
    }
  }

  fields_assign() {
    if (_LOOKUP0 == 1) {
      if (_VALID == 1) {
        ing_obj_bus.nexthop_index = nexthop_index;
        ing_obj_bus.myip_0 = myip[31:16];
        ing_obj_bus.myip_1 = myip[15:0];
        ing_cmd_bus.routed = routed;
        ing_cmd_bus.myflag = myflag;
        ing_cmd_bus.action_ctrl_0 = action_ctrl_0;
      } else { // miss
        ing_obj_bus.nexthop_index = 0;
        ing_obj_bus.myip_0 = 0;
        ing_obj_bus.myip_1 = 0;
        ing_cmd_bus.routed = 0;
        ing_cmd_bus.my_fib_drop = 1;
        ing_cmd_bus.myflag = 0;
        ing_cmd_bus.action_ctrl_0 = 0;
      }
    }
  }
}

// Action dispatch implemented in the FSL logic unit.
function ifsl_my_ip_host_process() {
  if (ing_cmd_bus.action_ctrl_0[1:1] == 1) {
    ing_cmd_bus.myflag = 1; 
  } else if (ing_cmd_bus.action_ctrl_0[2:2] == 1) {
    ing_cmd_bus.myflag = 0; 
  } else if (ing_cmd_bus.action_ctrl_0[3:3] == 1) {
    ing_obj_bus.myip_0 = 0x0A0A; 
    ing_obj_bus.myip_1 = 0x0A0A;
  }
}

function INGRESS_PROCESS() {
  my_ipv4_host.lookup(0);
  ifsl_my_ip_host_process();
}
\end{lstlisting}

\section{Supplementary Experimental Results}
\label{app:load}

\subsection{DPU Traffic Distribution}

\begin{figure*}[!b]
	\centering
    \begin{minipage}[t]{0.5\textwidth}{
    \subfigure[DPU 1]{
		\includegraphics[width=.48\textwidth]{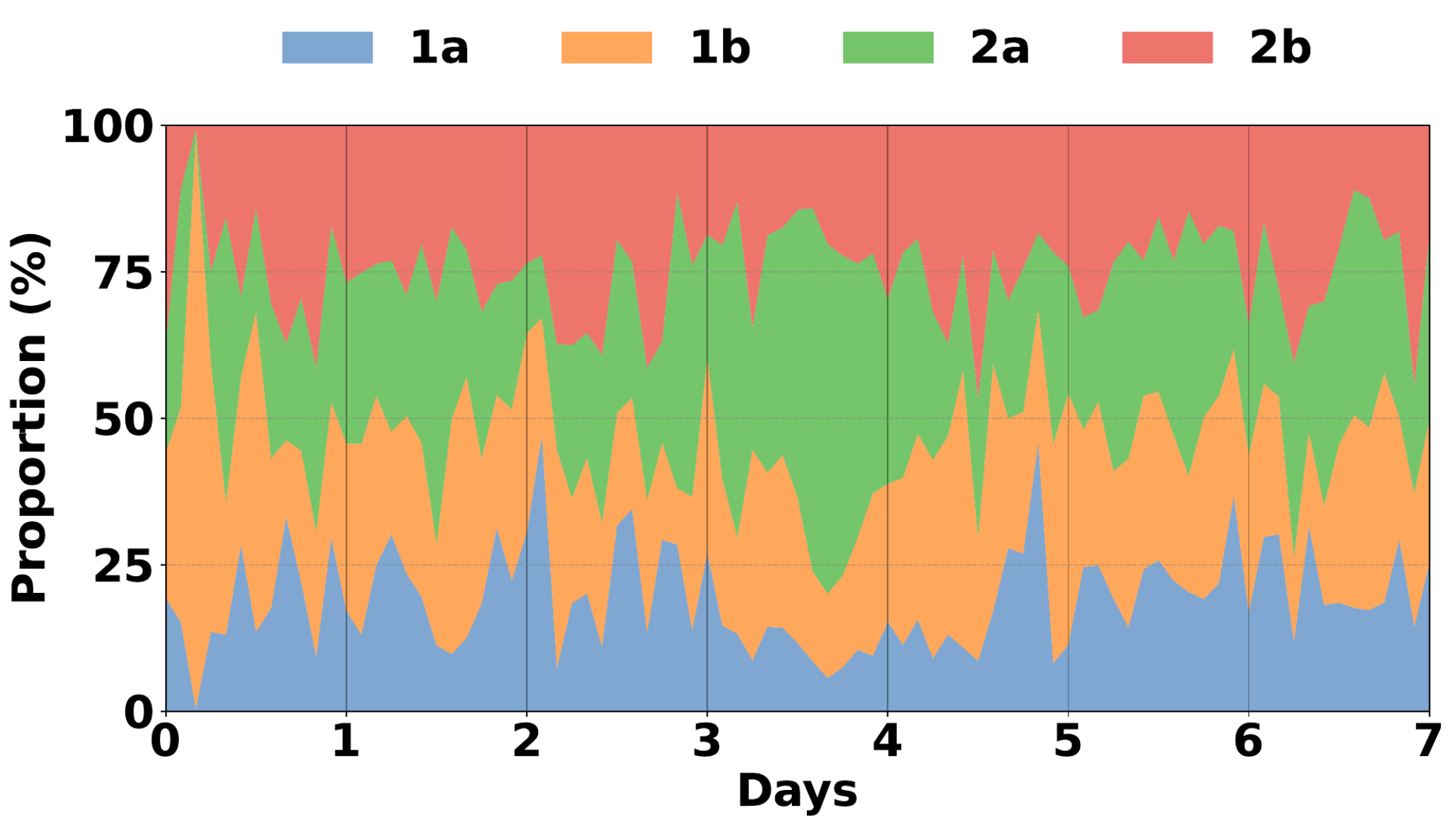}
        \vspace{-0.1in}
	}%
	\subfigure[DPU 2]{
		  \includegraphics[width=.48\textwidth]{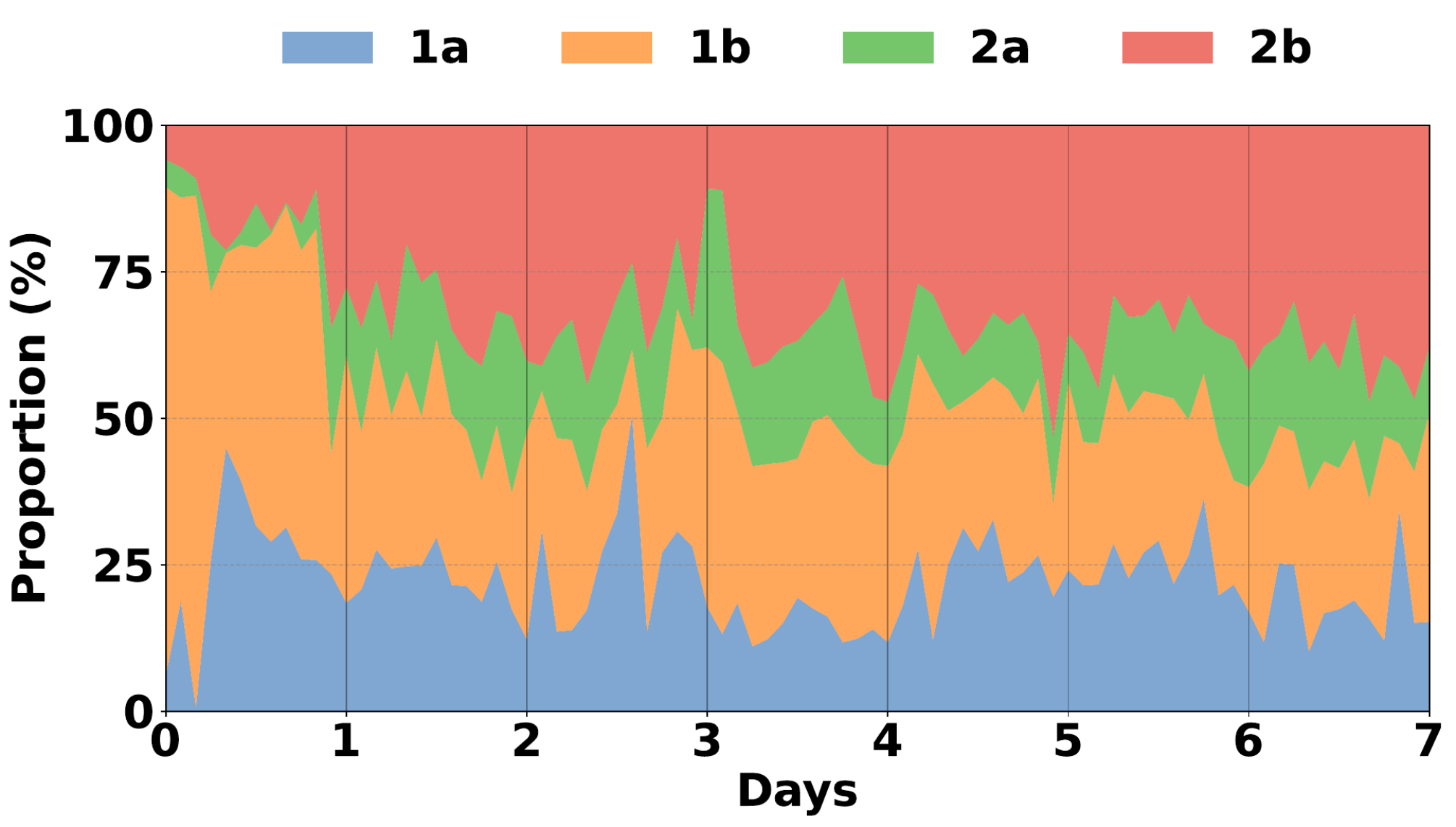}
          \vspace{-0.1in}
    }
    \caption{Intra-DPU traffic distribution.}
    \label{exp:supp:hash_port}
    }\end{minipage}%
    \begin{minipage}[t]{0.5\textwidth}{
    \subfigure[Node 1]{
		\includegraphics[width=.48\textwidth]{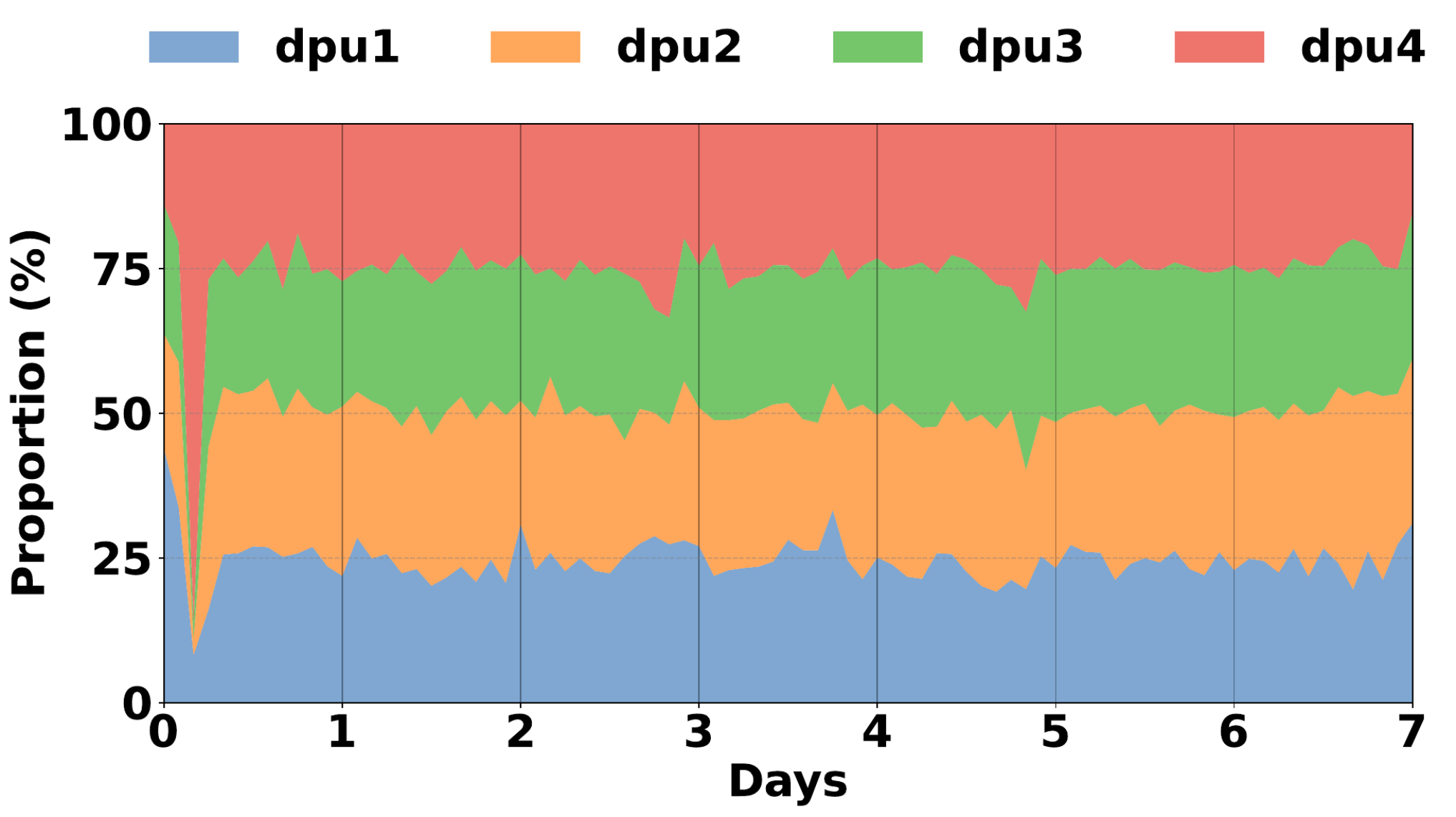}
        \vspace{-0.1in}
	}%
	\subfigure[Node 2]{
		  \includegraphics[width=.48\textwidth]{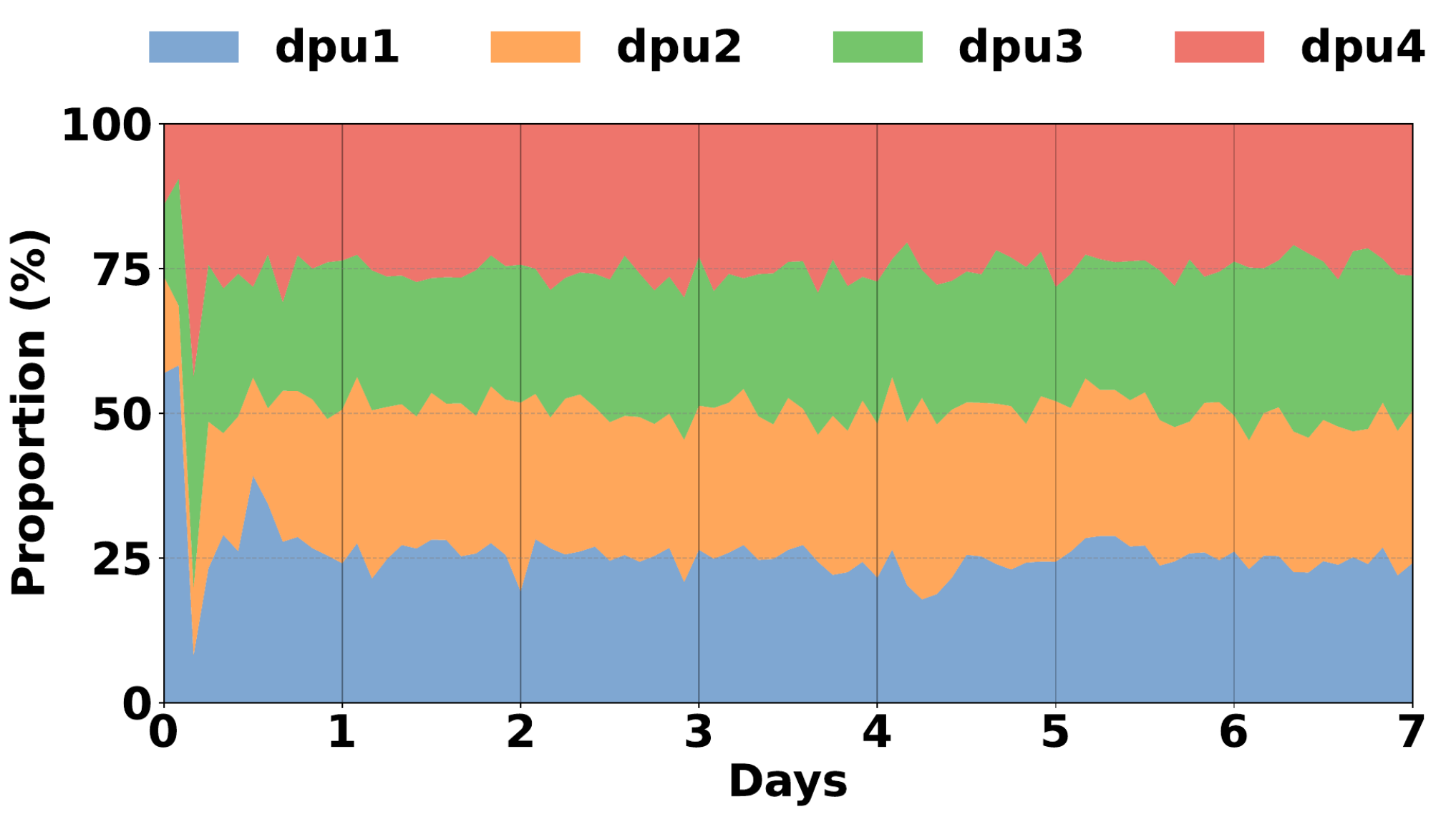}
          \vspace{-0.1in}
    }
    \caption{Inter-DPU traffic distribution.}
    \label{exp:supp:hash}
    }
	\end{minipage}%
\end{figure*}

We provide supplementary traffic distribution across different deployment instances in Figure \ref{exp:supp:hash_port} and \ref{exp:supp:hash}. 
These results demonstrate consistent characteristics: intra-DPU ports remain susceptible to imbalance and the traffic across DPUs is evenly distributed. 
This further validates the robustness and effectiveness of our dual-mode steering strategy across various production nodes.

\begin{figure*}[!b]
	\centering
	\subfigure[Pipeline 0]{
	\begin{minipage}[t]{0.5\textwidth}{
		  \includegraphics[width=1\textwidth]{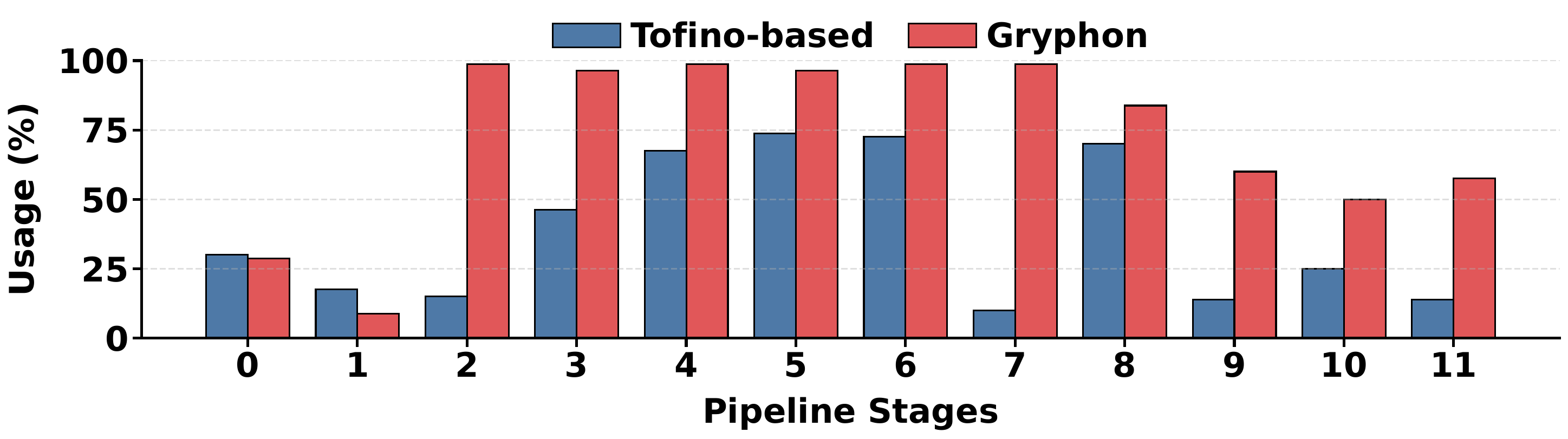}
		}
	\end{minipage}}%
	\subfigure[Pipeline 1]{
	\begin{minipage}[t]{0.5\textwidth}{
    	\includegraphics[width=1\textwidth]{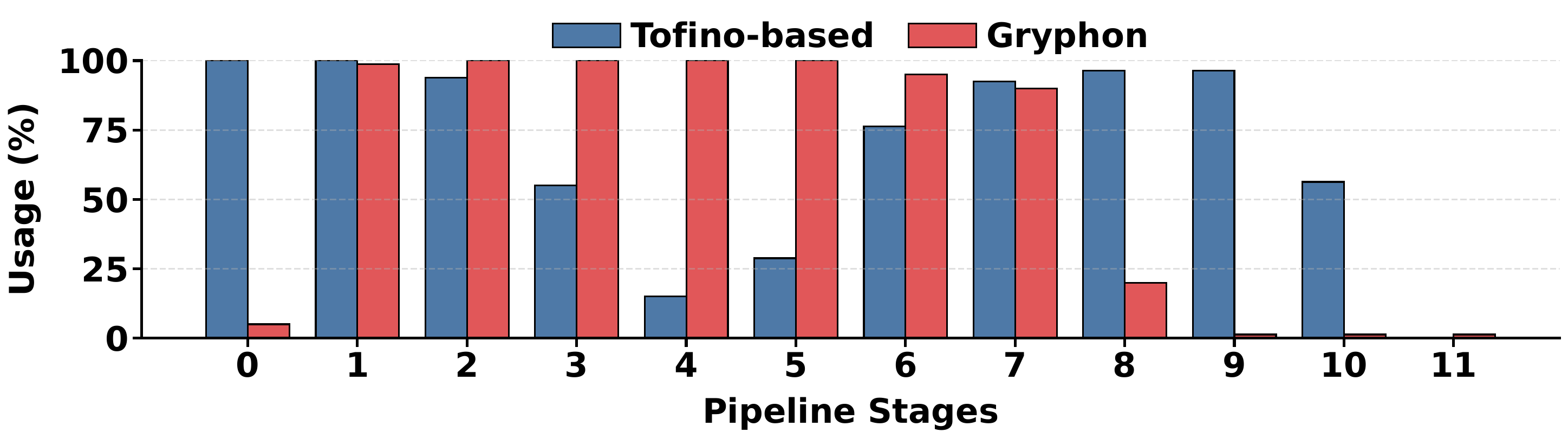}
		}
	\end{minipage}}\\
	\subfigure[Pipeline 2]{
	\begin{minipage}[t]{0.5\textwidth}{
		  \includegraphics[width=1\textwidth]{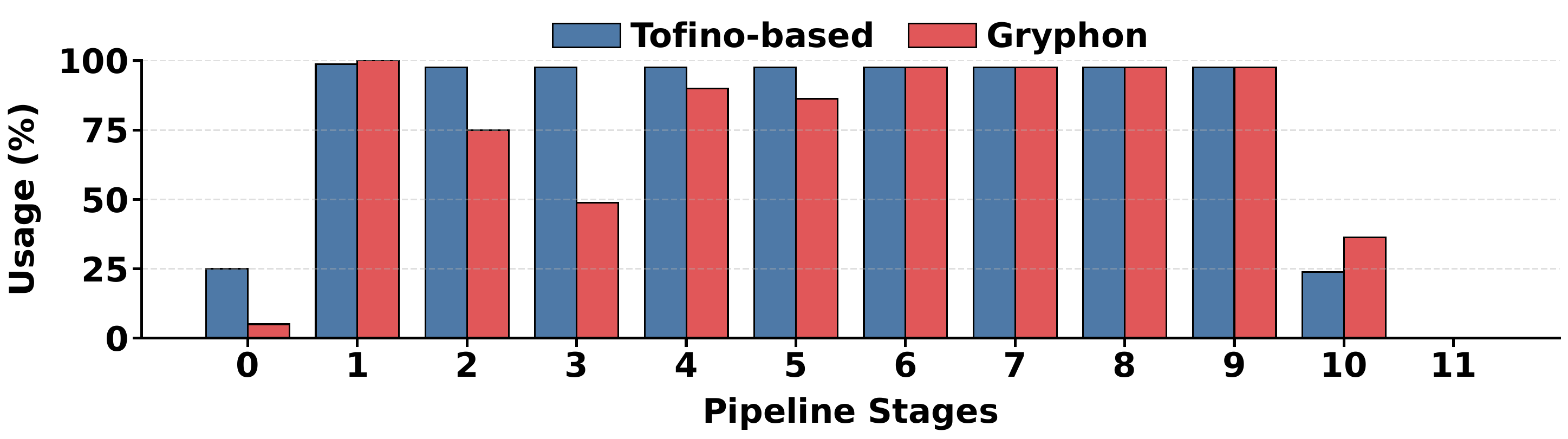}
		}
	\end{minipage}}%
	\subfigure[Pipeline 3]{
	\begin{minipage}[t]{0.5\textwidth}{
		  \includegraphics[width=1\textwidth]{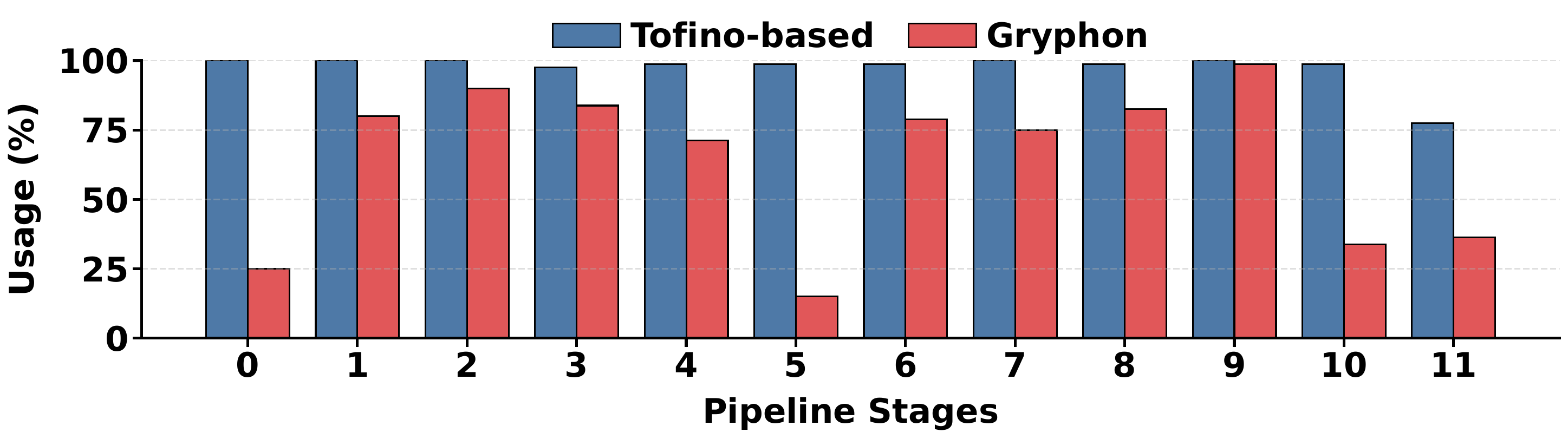}
		}
	\end{minipage}}\\
	\caption{SRAM usage over stages across pipelines.}
    \label{exp:sram}
    \vspace{-0.1in}
\end{figure*}

\subsection{Detailed SRAM Usage}

We present the detailed per-stage SRAM usage for each pipeline in Figure \ref{exp:sram}.
The results show that the Tofino-based gateway exhausts nearly all SRAM capacity across almost every stage in Pipeline 3, creating a rigid resource bottleneck that severely limits further scalability.